\documentclass{IEEEtran}
\usepackage{graphicx}
\usepackage{amsmath}
\usepackage{cite}
\usepackage{booktabs}
\usepackage{subfigure}

\title{A Deep Neural Network Deployment Based on Resistive Memory Accelerator Simulation}
\author{Tejaswanth Maram, Ria Barnwal, Dr. Bindu B}
\begin{document}
\maketitle

\begin{abstract}
The objective of this study is to illustrate the process of training a Deep Neural Network (DNN) within a Resistive RAM (ReRAM) Crossbar-based simulation environment using CrossSim, an Application Programming Interface (API) developed for this purpose. The CrossSim API is designed to simulate neural networks while taking into account factors that may affect the accuracy of solutions during training on non-linear and noisy ReRAM devices. ReRAM-based neural cores that serve as memory accelerators for digital cores on a chip can significantly reduce energy consumption by minimizing data transfers between the processor and SRAM and DRAM. CrossSim employs lookup tables obtained from experimentally derived datasets of real fabricated ReRAM devices to digitally reproduce noisy weight updates to the neural network. The CrossSim directory comprises eight device configurations that operate at different temperatures and are made of various materials. This study aims to analyze the results of training a Neural Network on the Breast Cancer Wisconsin (Diagnostic) dataset using CrossSim, plotting the innercore weight updates and average training and validation loss to investigate the outcomes of all the devices.
\end{abstract}

\section{Introduction}
When we talk about the existing memory technologies, we mainly talk about SRAM, DRAM and flash memory. While all of them are charge based storage devices, they are used in different places according to their varied performances in different areas. SRAM devices are high speed, low power consuming devices which are often used as small cache memory units in the computers because of it’s complexity to scale in size and costlier to manufacture. DRAM being a cheaper technology and better scalable, it is used for bigger RAM units which stores the temporary data that is required by the computer for the current program at volatile level. Flash memory is a highly scalable and low-power-consuming non-volatile memory technology, which makes it an ideal candidate for use as a secondary storage unit in computer systems.

\begin{figure}[htbp]
    \centering
    \includegraphics[width=0.4\textwidth]{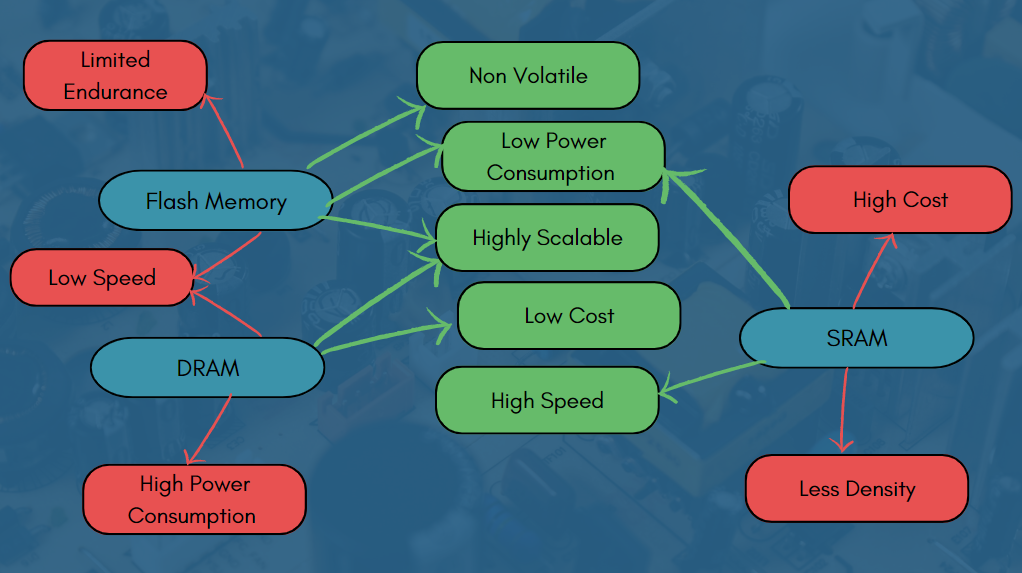}
    \caption{Existing memory technologies and their properties.}
    \label{fig1}
\end{figure}

It is desired by the industry to create an ideal memory technology which has most of the upsides and very few if not none of the downsides of the current memory technologies. Hence with the aim of reinventing the current memory storage techniques, the industry is actively researching the development of a new hierarchy of memory often known as emerging memory technologies. The purpose of these emerging memory technologies is to enable the strengths of existing memory technologies like the rapid toggling capability of SRAMs, high storage capacity equivalent to DRAMs, and non-volatile capability of Flash memory. thus having the potential to be highly appealing alternatives to the next-generation memory hierarchy. \cite{article2}\\
In contrast to charge-based memories technologies, these memory technologies works on the principle of making changes in the resistance of the cells to store the information. Few examples of this emerging memory technologies are :  (i) PCM or phase change memory, (ii) STT-MRAM or spin-transfer torque magnetoresistive random access memory, and (iii) ReRAM or resistive random access memory.
ReRAM is characterized as a non-volatile memory type that has the ability to preserve data even during the power absence. ReRAM stores the data in form of Low and High resistive states. The Metal Insulator Metal (MIM) structure of this memory type allows the fabrication to be simple. It comprises an insulating layer (I) positioned between two metal (M) electrodes. The storage and retrieval of data is dependent on the creation and breaking of a conductive filament between the electrodes, resulting in the low resistance state (LRS) and high resistance state (HRS) respectively. These states will be retained until it is changed with a writing voltage thus making the ReRAMS non volatile.

\begin{figure}[htbp]
    \centering
    \includegraphics[width=0.35\textwidth]{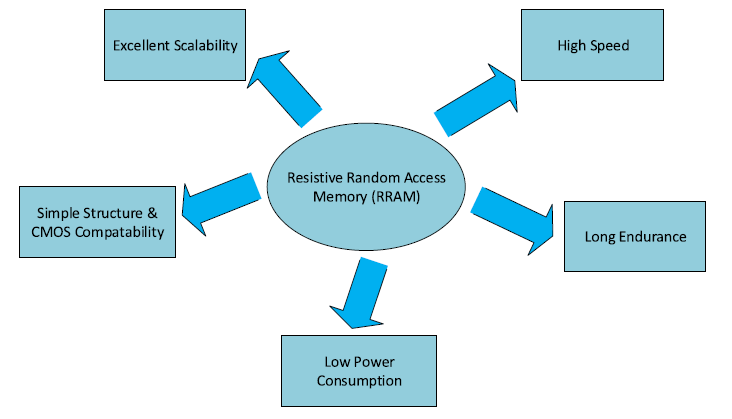}
    \caption{Advantages of ReRAM \cite{article3}}
    \label{fig2}
\end{figure}

A ReRAM uses a set voltage (Vs) to form a conductive filament between the electrodes representing the Low Resistive State(LRS). And a reset voltage (Vr) to rupture the filament representing the High Reistive State (HRS).

\begin{figure}[htbp]
    \centering
    \includegraphics[width=0.4\textwidth]{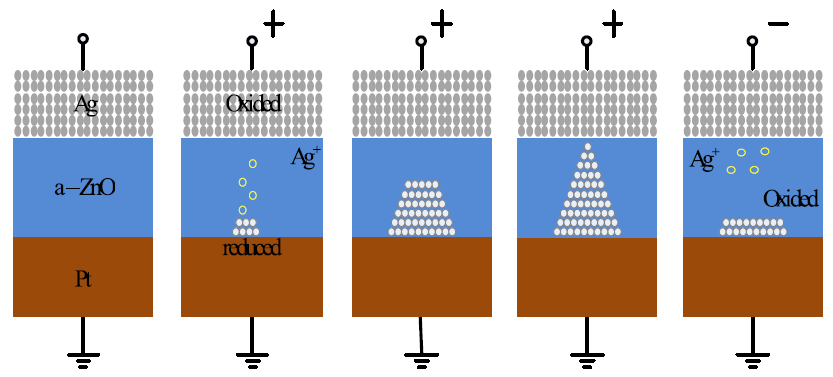}
    \caption{Switching mechanism of ReRAM \cite{article3}}
    \label{Fig3}
\end{figure}

In comparison with a conventional CPU, resistive memory crossbars can reduce the energy required to perform computations in neural algorithms by five orders of magnitude or more. Data movement between the processor, SRAM (static random access memory), and DRAM (dynamic random access memory) dominates computational energy in data intensive applications. Memristor and resistive crossbars can effectively handle large data sets by minimizing data movement, leveraging analog operations, and increasing memory density on a single chip. This newer approach of computing with analog memory based neural cores acting as an accelerator to the digital core will make the data intensive applications much more power efficient, compact and faster. \\
Resistive memory devices are essentially two terminal programmable resistors. The device's resistive state has no impact with a low input voltage and can essentially be used to read the resistance of the memory cell. While as to write a value to the device, a higher voltage is applied for a set period of time which will be stored as an analog value of resistance in the memory cell. When a crossbar of resistive memory cells is applied as a neural network, the resistance of each cell behaves as a weight simulating a neural synaptic connection. As a result, neuromorphic systems utilizing such tools have gained popularity. In theory, the resistive memory would react in a linearly predictable and controllable manner, enabling them to be set to any given analog value.

The use of resistive memory crossbar has the potential to accelerate two important operations in many neural algorithms: parallel read (or vector matrix multiply) and parallel write (or rank one outer product). Sparse coding, restricted Boltzmann machines, and backpropagation are examples of neural algorithms that depend significantly on these two operations. The processing of inputs and outputs on a crossbar varies based on the specific algorithm used.

\begin{figure}[htbp]
    \centering
    \includegraphics[width=0.32\textwidth]{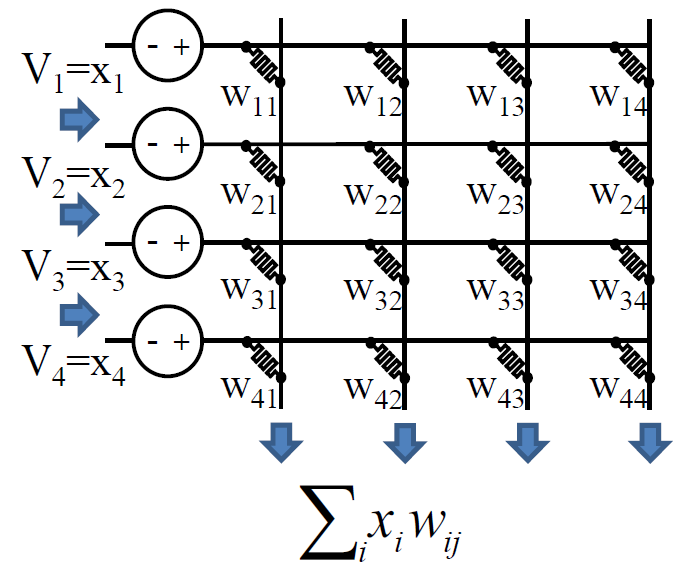}
    \caption{A parallel read (or) vector matrix multiply on crossbar array \cite{article1}}
    \label{Fig4}
\end{figure}

\begin{figure}[htbp]
    \centering
    \includegraphics[width=0.32\textwidth]{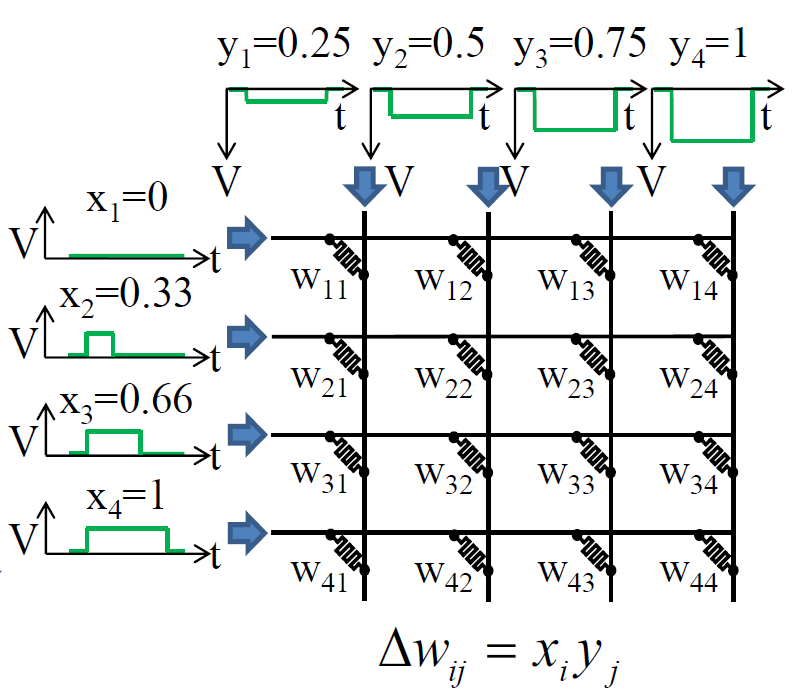}
    \caption{A parallel write (or) rank one outerproduct \cite{article1}}
    \label{Fig5}
\end{figure}

\section{Materials and Methodology}
The Breast Cancer Wisconsin (Diagnostic) Data Set \cite{Dua:2019} was utilized the DNN for this paper. This dataset includes 569 instances of features derived from samples obtained by inserting a thin needle into a breast mass and extracting cells called FNA for analysis. The features record the characteristics of cell nuclei in the images. The dataset includes the mean, standard error, and mean of the three largest values for 10 attributes computed for each image, resulting in a total of 30 features. These attributes include radius (the average distance from the center to points on the perimeter), texture (the standard deviation of gray-scale values), perimeter, area, smoothness (the local variation in radius lengths), compactness (square of the perimeter, divided by area, minus 1.0), concavity (the severity of concave portions of the contour), concave points (the number of concave portions of the contour), symmetry, and fractal dimension (which approximates the coastline). A deep neural net is to be trained on these 569 instances of 30 features to classify the given set of features into either one of the two classes: Malignant and Benign. \\

Exploratory data analysis will give us a summary of the main characteristics of the dataset in the form of visual designs, graphs, etc. To check the balance of the dataset we visualized the count of each class as a seaborn count plot. And, The correlation between the attributes can be visualized as a heatmap.

\begin{figure}[htbp]
    \centering
    \includegraphics[width=0.37\textwidth]{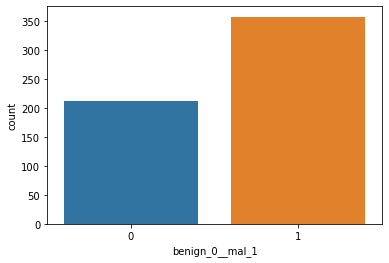}
    \caption{Count plot of the two classes benign and malignant}
    \label{Fig6}
\end{figure}

\begin{figure}[htbp]
    \centering
    \includegraphics[width=0.31\textwidth]{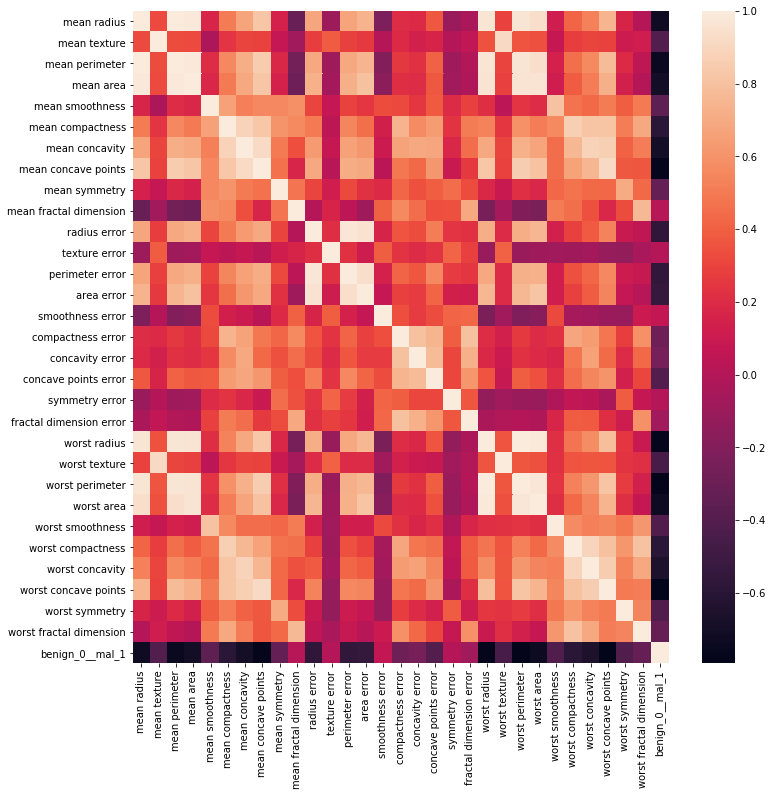}
    \caption{Heatmap representing the correlation between the attributes}
    \label{Fig7}
\end{figure}

To simulate this famous dataset on ReRAM crossbars, a tool called CrossSim has been used for modeling analog in-memory computing for neural networks and linear algebra operations. It is designed to simulate the impact of analog hardware effects in resistive crossbars on the algorithm's accuracy. The simulator takes into account various factors that affect the accuracy of training a neural network, such as the non-linear conductance update, write symmetry, write stochasticity, and device-to-device variability. CrossSim generates lookup tables from experimental data of a ReRAM device to digitally simulate these factors. Its accuracy simulation and co-design capabilities make it a valuable tool for designing and optimizing analog hardware for neural network applications. The details of this tool and its applications are discussed in a published research paper \cite{article1}.

\begin{figure}[ht]
  \centering
  \subfigure[Decreasing Conductance vs CDF]{\label{Fig8a}\includegraphics[width=0.22\textwidth]{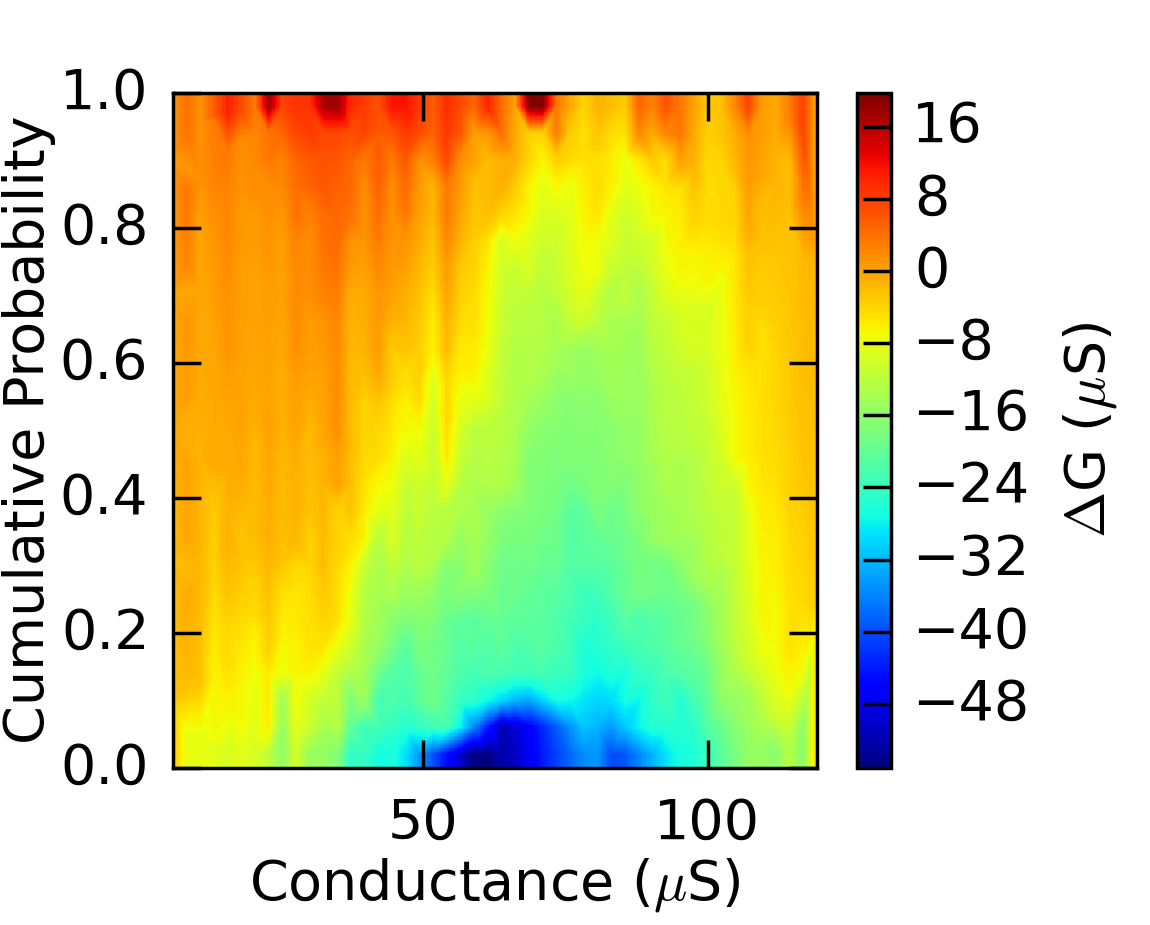}}
  \hfill
  \subfigure[Decreasing Conductance vs Delta G]{\label{Fig8b}\includegraphics[width=0.22\textwidth]{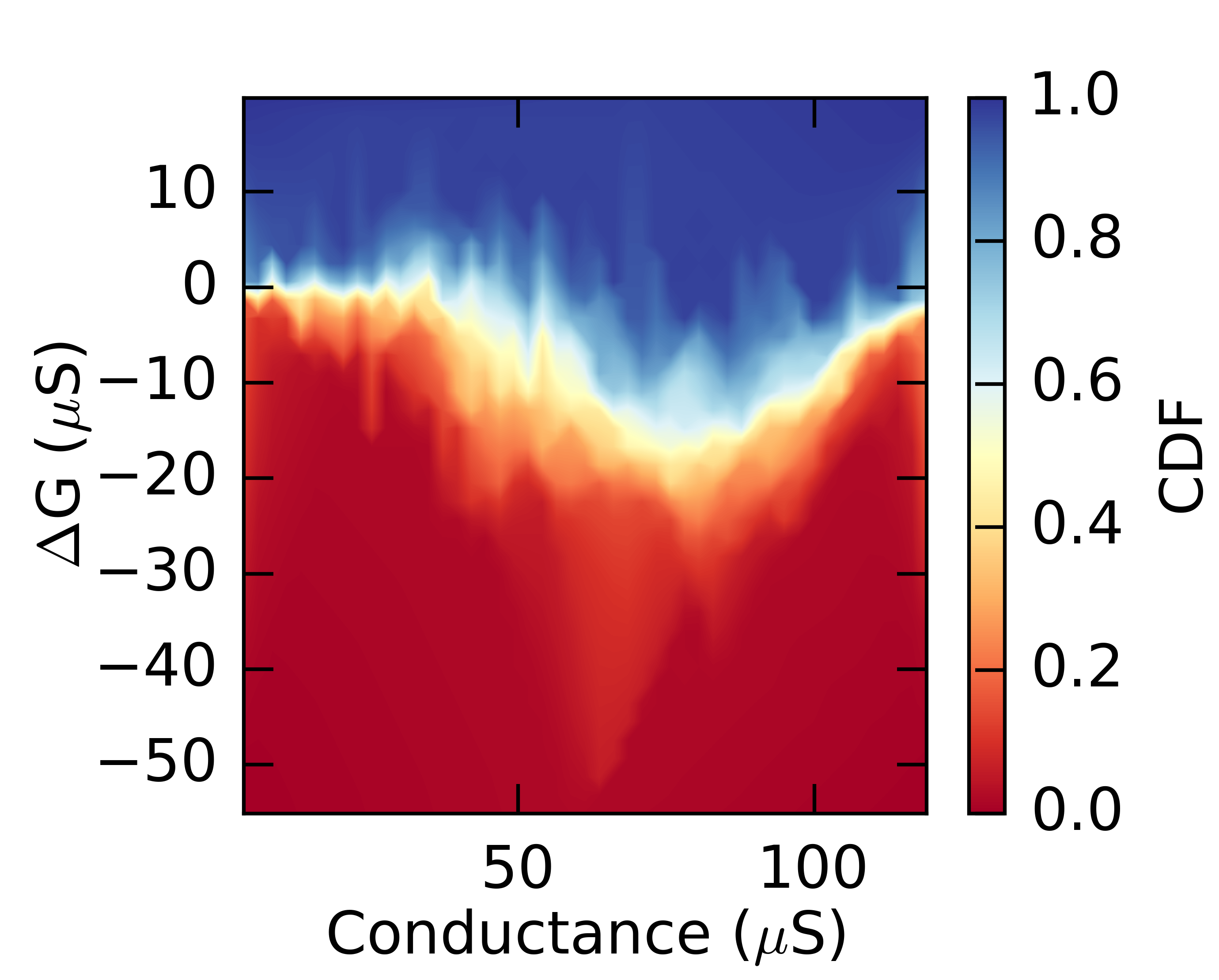}}
  
  \medskip
  
  \subfigure[Increasing Conductance vs CDF]{\label{Fig8c}\includegraphics[width=0.22\textwidth]{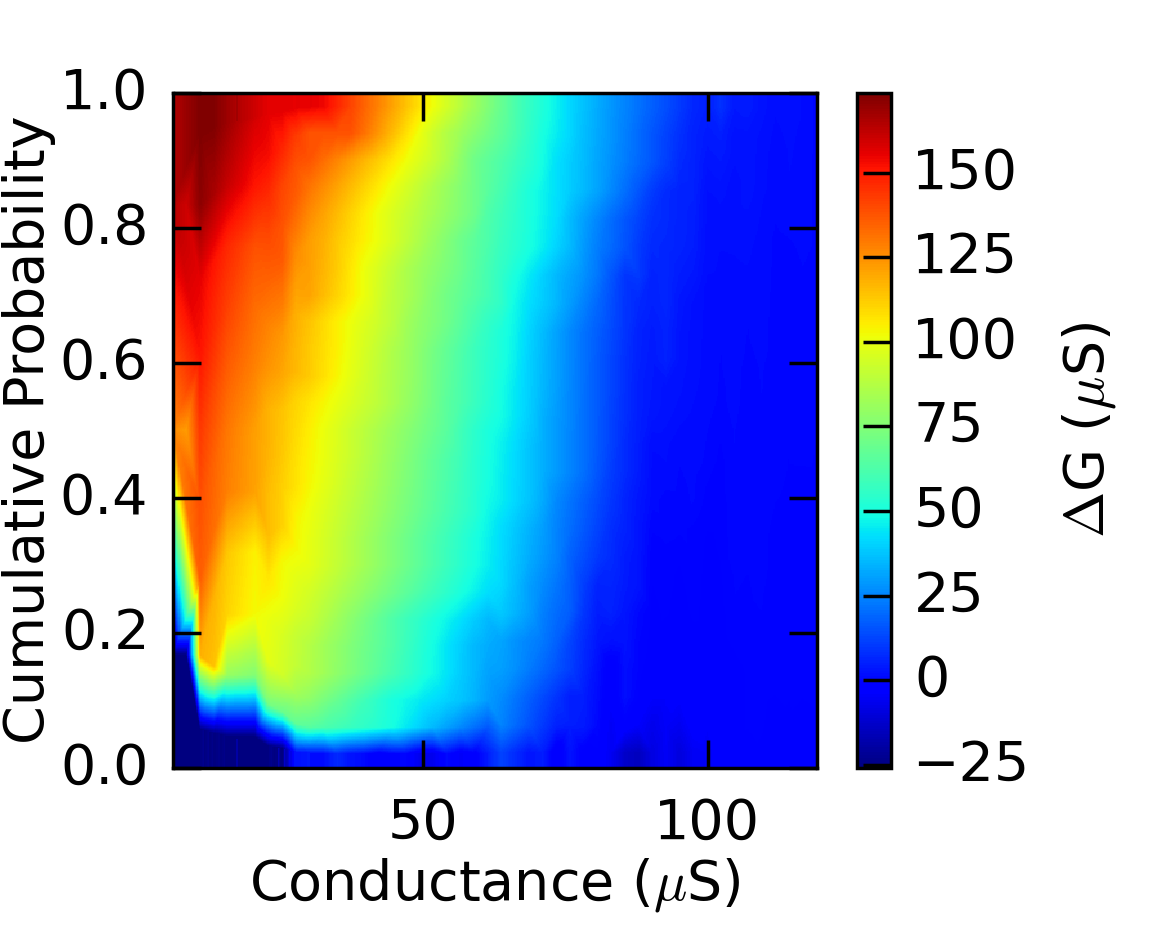}}
  \hfill
  \subfigure[Increasing Conductance vs Delta G]{\label{Fig8d}\includegraphics[width=0.22\textwidth]{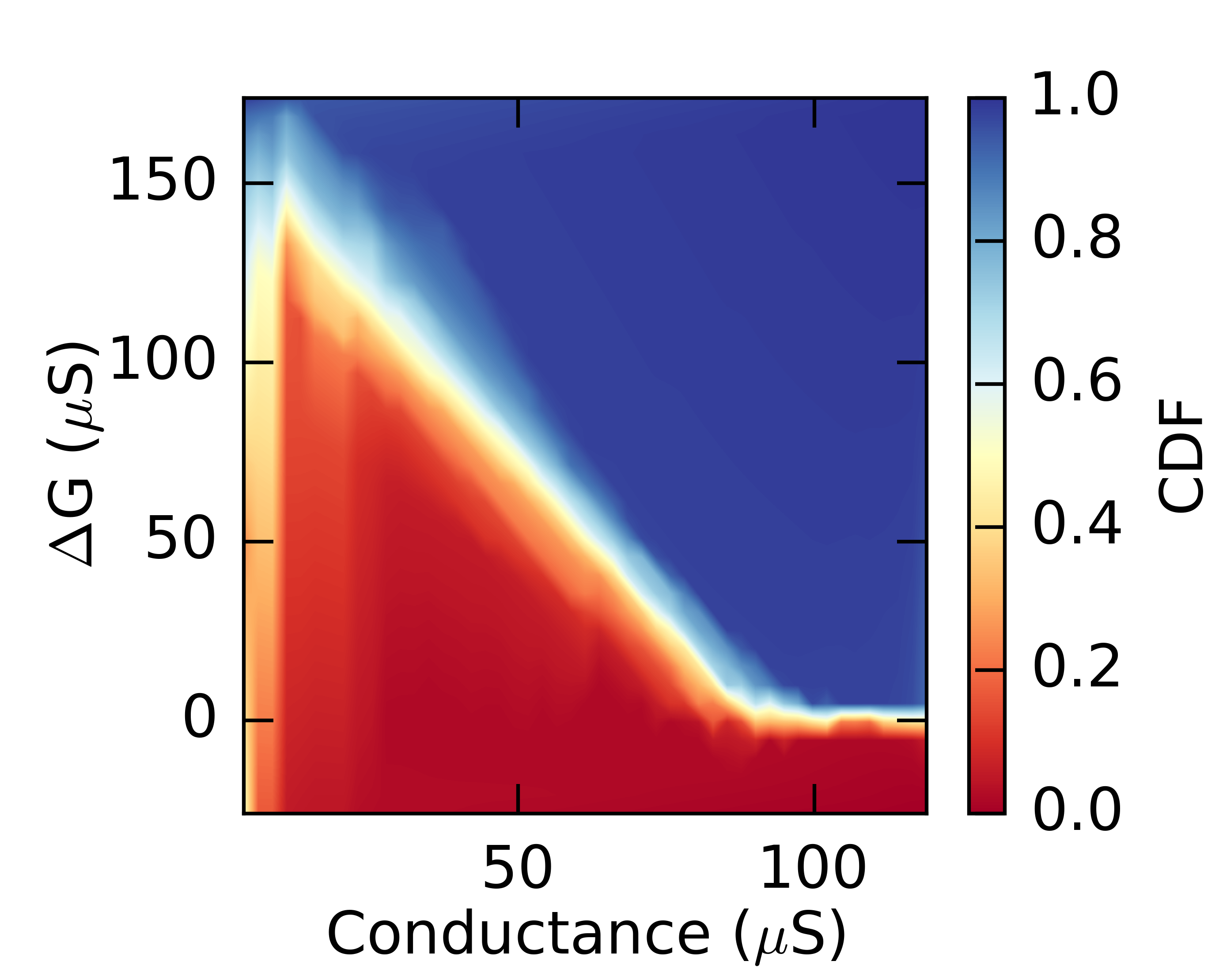}}
  
  \caption{Conductance plots for TaOx Device}
  \label{fig:abcd}
\end{figure}

CrossSim GitHub repository is provided with more than 180 device configurations of various materials, mechanisms ,and temperature conditions. The experimental data of the devices used to generate the lookup tables can be visualized as conductance plots. Figures 8a, 8b, 8c ,and 8d represent the conductance plots of TaOx ReRAM devices fabricated at Sandia, ca. 2016 \cite{8720596}

This paper utilizes the broadly classified 10 configurations of devices available on the CrossSim which are: DWMTJ SOT 0K, DWMTJ SOT 300K, DWMTJ SOT 400K, DWMTJ STT 0K, DWMTJ STT 300K, DWMTJ STT 400K, ENODe, and TaOx.

A simple deep neural network is built to train the dataset simulated on the above 8 device configurations. The simulation and model parameters of the CrossSim are set by default. The DNN consists of 3 layers, first layer is the input layer with 30 neurons for the input of 30 features. The input layer uses the sigmoid function for the activation purpose. The second layer is the hidden layer with 15 neurons and follows the sigmoid function as the activation purpose. The nonlinearity of the sigmoid function can capture the complex relationship between the input and the output variables. The “S” – shaped curve bounded between the range of 0 and 1 will be very sensitive to small changes. Figure 10 represents the logistic curve which is an example of the sigmoid function on the XY-axis graph.

\begin{figure}[htbp]
    \centering
    \includegraphics[width=0.35\textwidth]{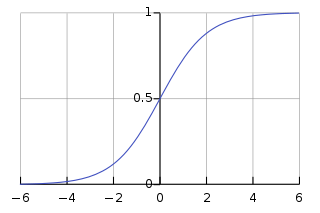}
    \caption{logistic function \cite{10.1007/3-540-59497-3_175}}
    \label{Fig9}
\end{figure}

The third layer of the DNN is the output layer consisting of 2 nodes for the binary classification of the output into either of the two classes namely Malignant and Benign. The output layer uses the SoftMax function as the activation function. By applying the SoftMax function, a set of n actual numbers is converted into n real numbers whose sum is equal to 1. The neural net then turns on the node with higher probability thus classifying the given input into either of the 2 classes. Following are the equations of the sigmoid and the SoftMax functions respectively.

Sigmoid Function:
\begin{equation}
    S(x) = \frac{1}{1+e^{-x}} = \frac{e^x}{e^x+1} = 1-S(-x)
\end{equation}

Softmax Function:
\begin{equation}
    \sigma (z)_i = \frac{e^{z_i}}{\sum_{j=1}^{K} e^{z_j}}  \qquad \text{for } i = 1,\dots,K
\end{equation}

During the backpropagation of the network, the network uses a cost function C(w) to calculate the error in the prediction and backpropagate to adjust the weights of the neurons in order to minimize the error. One widely used cost function is the quadratic cost function. The neural net applies a gradient descent algorithm to the cost function C(w) and finds the minima which would result in the least value of error possible in the prediction made by the DNN. The step of the gradient also plays an important role in the performance of a neural network. A smaller step value would result in higher time consumption to reach the minima while a bigger step value could result in inaccurate minima. An optimizer algorithm would help find the right step size for the gradient descent algorithm and one such algorithm that is used is Adam’s algorithm. Figure 10 shows the performance of Adam algorithm on compared to other famous optimization algorithms.

\begin{figure}[htbp]
    \centering
    \includegraphics[width=0.35\textwidth]{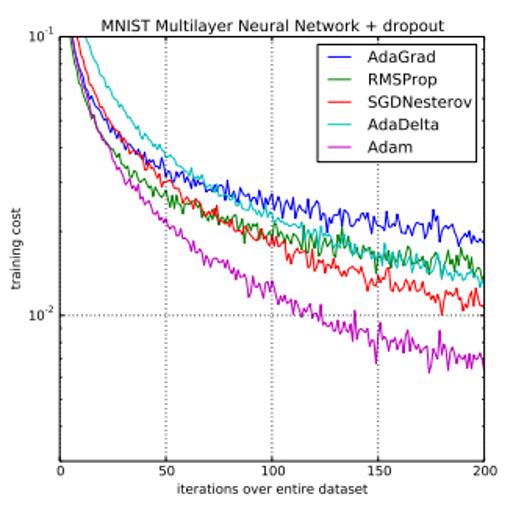}
    \caption{Adam algorithm performance comparison on the MNIST dataset \cite{article5}}
    \label{Fig10}
\end{figure}

Quadratic Cost Function:
\begin{equation}
    \sigma (z)_i = \frac{1}{2n} \sum_x {[y(x)-a^L(x)]}^2
\end{equation}

\section{Results and Discussion}
The neural net with the above mentioned architecture and parameters has been trained and validated on the 8 multiple device configurations. The dataset was partitioned into three subsets, with 40\% allocated for training, 30\% for validation, and 30\% for testing purposes. The observations in the innercore update errors have been visualized on Target update vs Real update plot and a update error vs probability density plot. The plots have been mapped for the numeric, standard and multi update configurations\\
The training was done for 80 epochs on each device. The training curves are plotted against the average loss in the predictions made for the training data and the average loss in the predictions made for the validation data. The plot for training loss starts from the epoch 0 while the plot for validation loss starts from epoch 1 for the better scaling of the graph pictures. The innercore updates plots shows the weight updates happened in each epoch on a scatter plot. The plot displays the real update vs target update in the inner core and also the update error and its probability density.\\
The DWMTJ: domain-wall magnetic tunnel junction is a spintronic device that stores data inform of magnetic spins. This device has been simulated with the data in three different temperatures: 0K, 300K and 400K, and in two different DW propagation mechanisms: spin transfer torque (STT) and spin orbit torque (SOT) \cite{osti_1784862}. The ENODe device refers to the electrochemical random access memory (ECRAM). The multi model of ENODe uses lookup table based on 9 devices\cite{osti_1775154}. The TaOx device uses the lookup table from the experimental data generated by the TaOx ReRAM devices fabricated at Sandia. The multi model for TaOx listed uses the medium set from CrossSim \cite{8720596}.\\
Observations show that all the devices perform with accuracies well above 92\%. It also can be clearly seen from the weight update plots of DWMTJ that the lesser temperatures will decrease the noise in the updates causing less error. The higher temperature induces more noise and error in the weight updates of a device. DWMTJ SOT at 400K has the highest accuracy of 97.8\%. ENODe displayed the least accuracy of 92.9\% in the standard and multi modes. \\

\begin{tabular}{|l|c|c|c|}
\hline
\textbf{Device Configuration} & \multicolumn{3}{|c|}{\textbf{Training Accuracy}} \\
\cline{2-4}
& \textbf{Numeric} & \textbf{Standard} & \textbf{Multi} \\
\hline
DWMTJ SOT 0K & 96.4\% & 97.8\% & 95.7\% \\
\hline
DWMTJ SOT 300k & 97.1\% & 97.1\% & 97.1\% \\
\hline
DWMTJ SOT 400k & 97.8\% & 97.8\% &97.8\% \\
\hline
DWMTJ STT 0k & 96.4\% & 97.8\% & 97.8\% \\
\hline
DWMTJ STT 300k & 97.1\% & 96.4\% & 95.7\% \\
\hline
DWMTJ STT 400k & 96.4\% & 97.1\% & 97.8\% \\
\hline
TaOx & 95.7\% & 96.4\% & 97.1\% \\
\hline
ENODe 8 & 96.4\% & 92.9\% & 92.9\% \\
\hline
\end{tabular}

\subsection{DWMTJ SOT 0k}
\begin{figure}[htbp]
    \centering
    \includegraphics[width=0.4\textwidth]{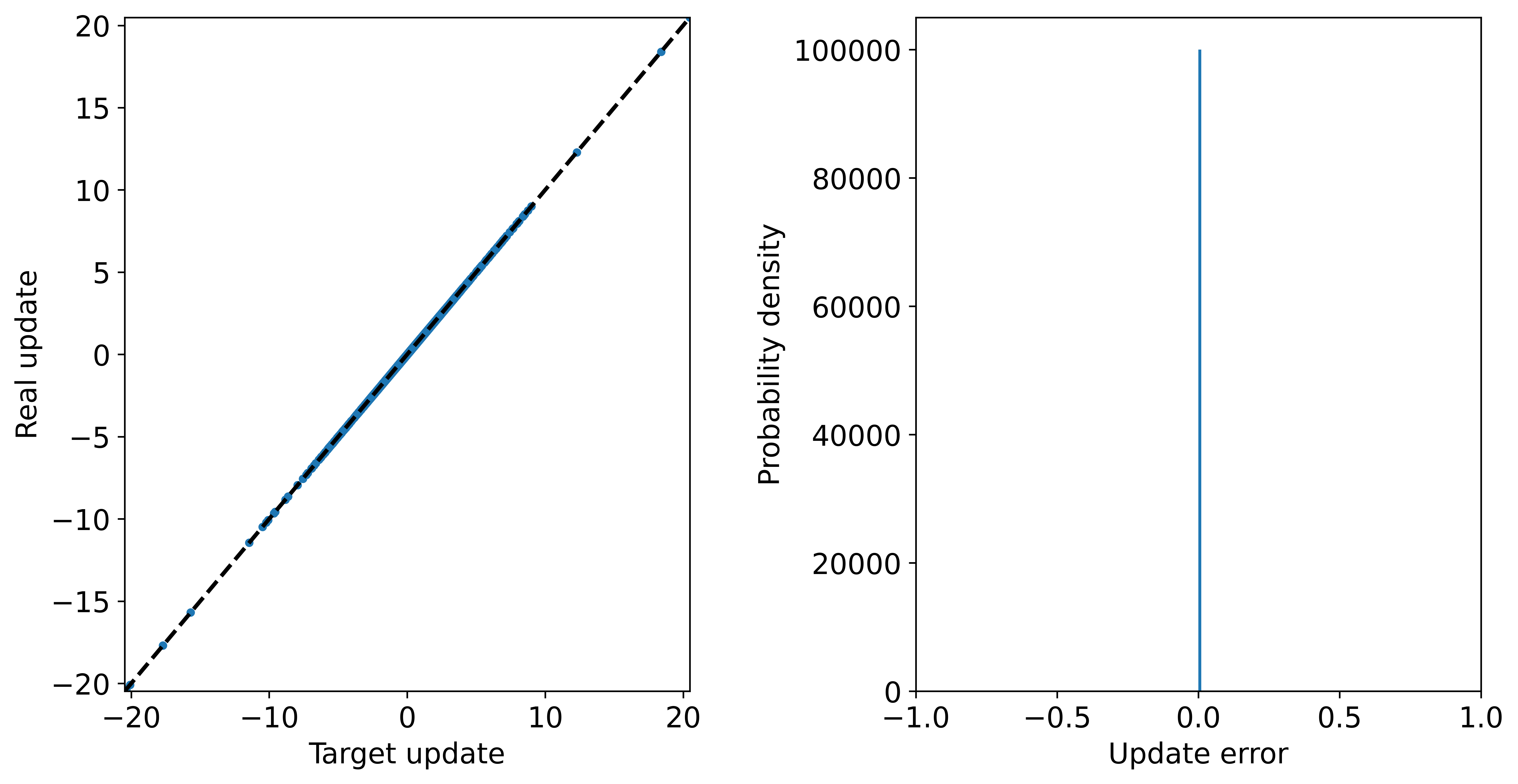}
    \caption{innercore update error Balanced DWMTJ SOT 0K numeric}
    \label{Fig11}
\end{figure}

\begin{figure}[htbp]
    \centering
    \includegraphics[width=0.32\textwidth]{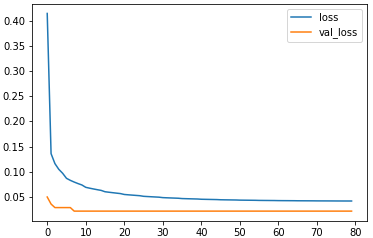}
    \caption{avg training loss vs avg validation loss}
    \label{Fig12}
\end{figure}

\begin{figure}[htbp]
    \centering
    \includegraphics[width=0.4\textwidth]{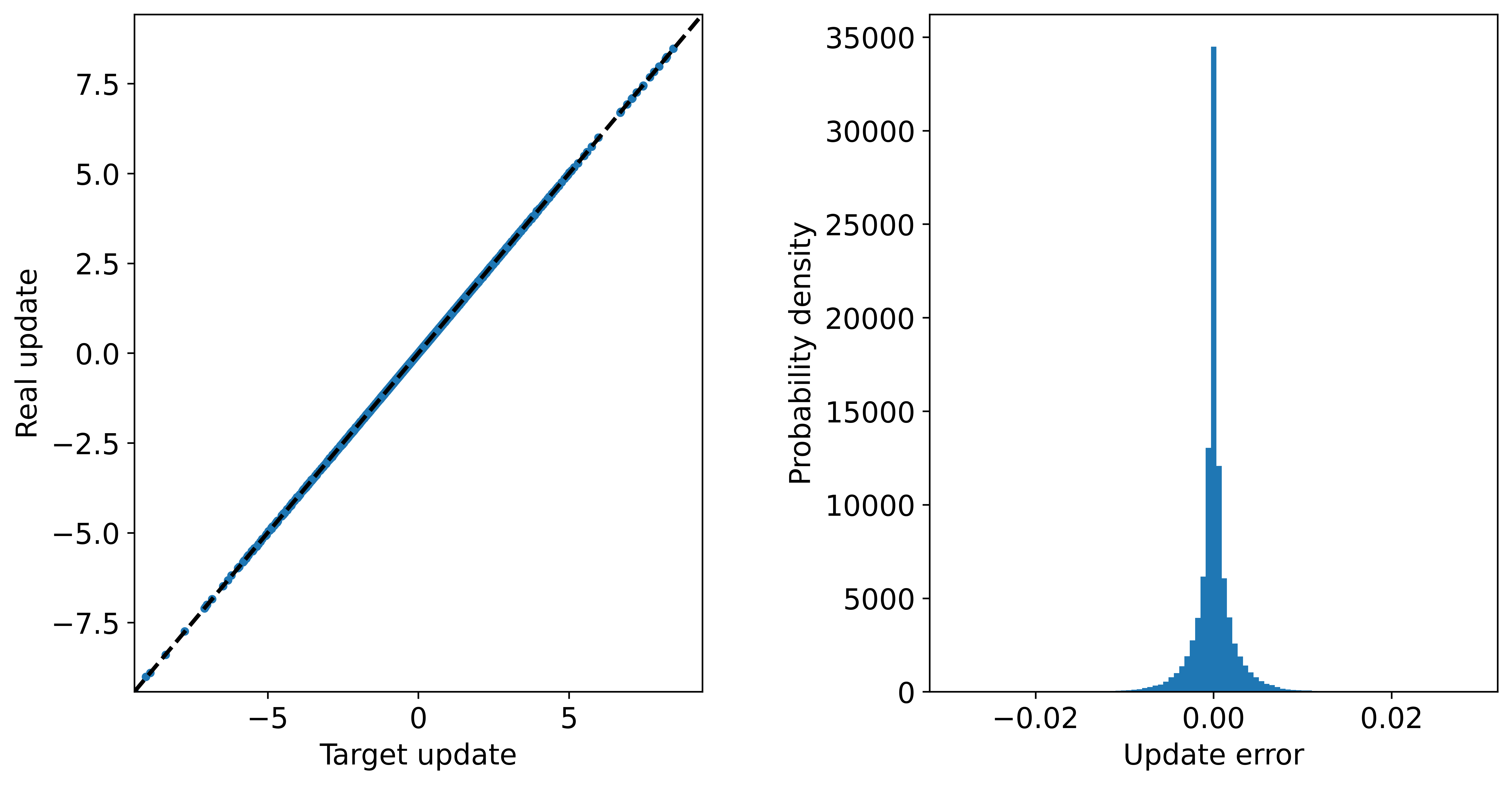}
    \caption{innercore update error Balanced DWMTJ SOT 0K standard}
    \label{Fig13}
\end{figure}

\begin{figure}[htbp]
    \centering
    \includegraphics[width=0.4\textwidth]{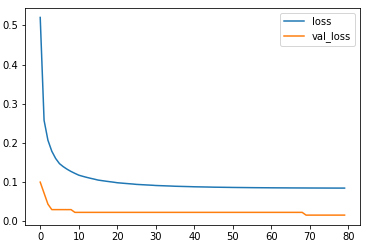}
    \caption{avg training loss vs avg validation loss}
    \label{Fig14}
\end{figure}

\begin{figure}[htbp]
    \centering
    \includegraphics[width=0.4\textwidth]{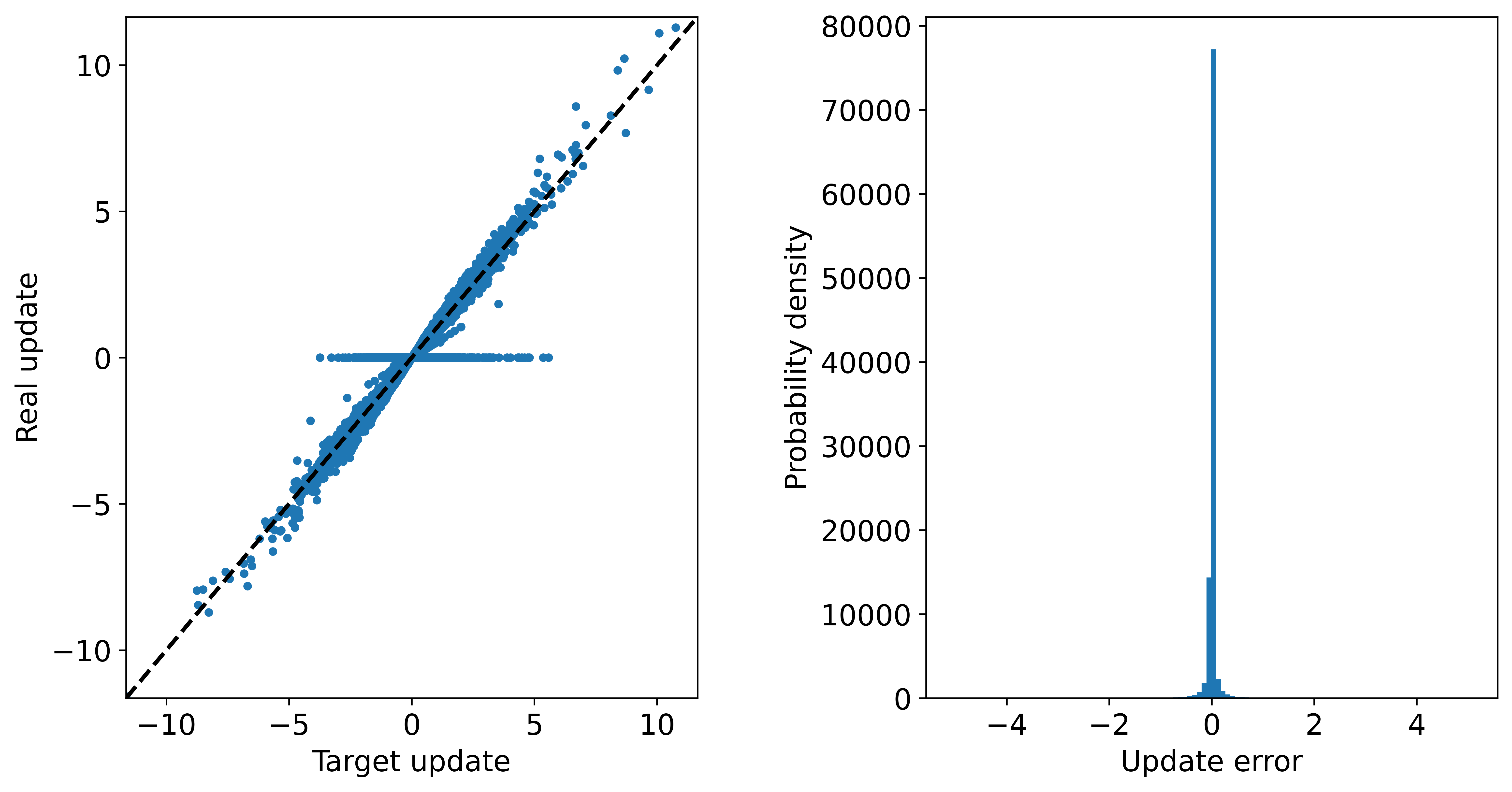}
    \caption{innercore update error Balanced DWMTJ SOT 0K multi}
    \label{Fig15}
\end{figure}

\begin{figure}[htbp]
    \centering
    \includegraphics[width=0.4\textwidth]{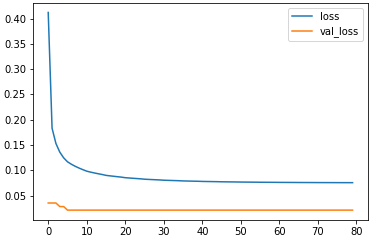}
    \caption{avg training loss vs avg validation loss}
    \label{Fig16}
\end{figure}

\subsection{DWMTJ SOT 300k}
\begin{figure}[htbp]
    \centering
    \includegraphics[width=0.4\textwidth]{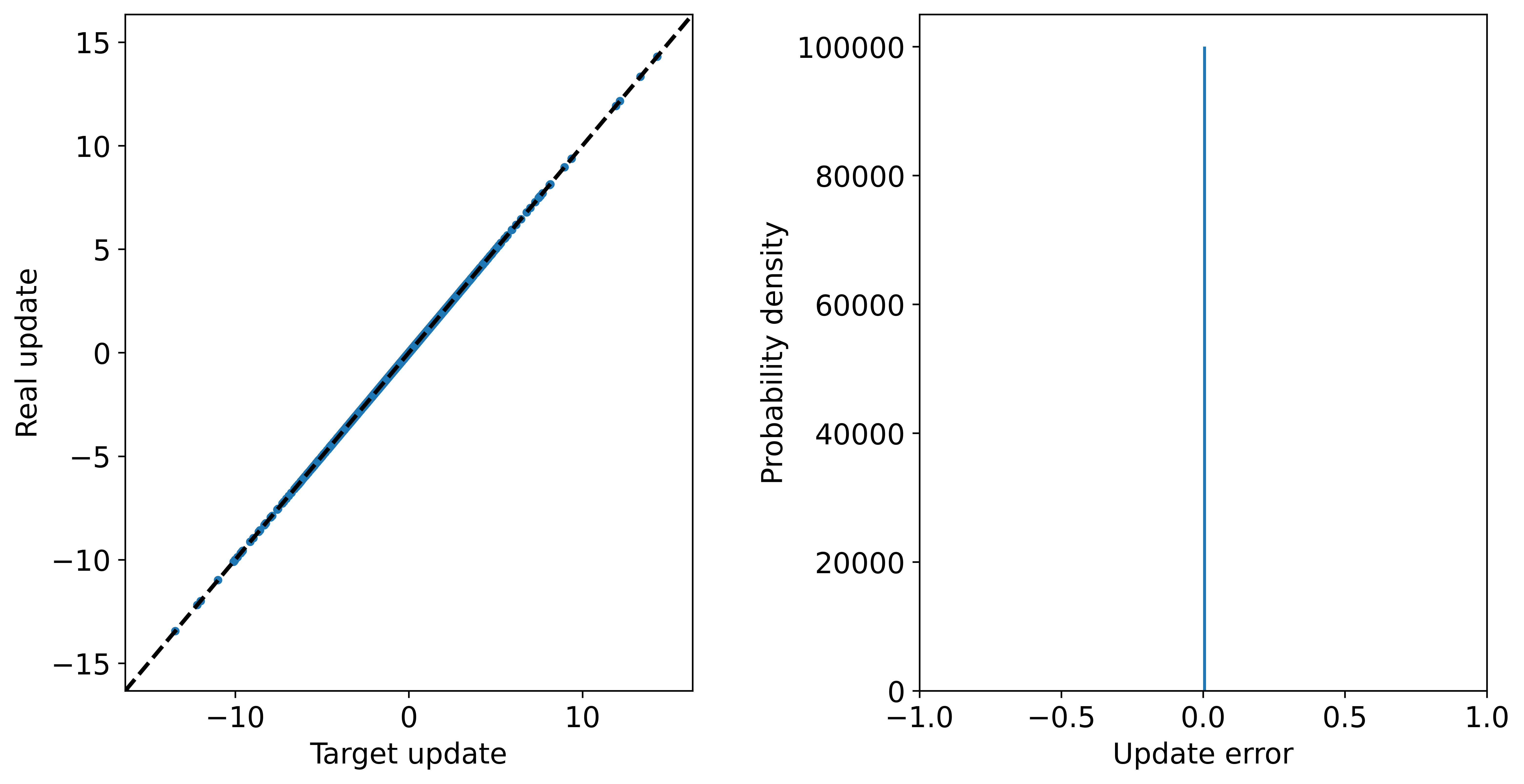}
    \caption{innercore update error Balanced DWMTJ SOT 300K numeric}
    \label{Fig17}
\end{figure}

\begin{figure}[htbp]
    \centering
    \includegraphics[width=0.37\textwidth]{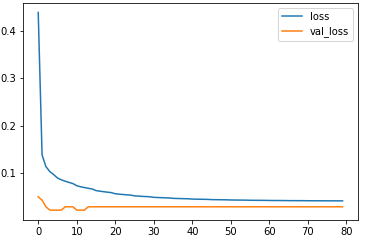}
    \caption{avg training loss vs avg validation loss}
    \label{Fig18}
\end{figure}
\vspace{10pt}
\vspace{10pt}
\begin{figure}[htbp]
    \centering
    \includegraphics[width=0.39\textwidth]{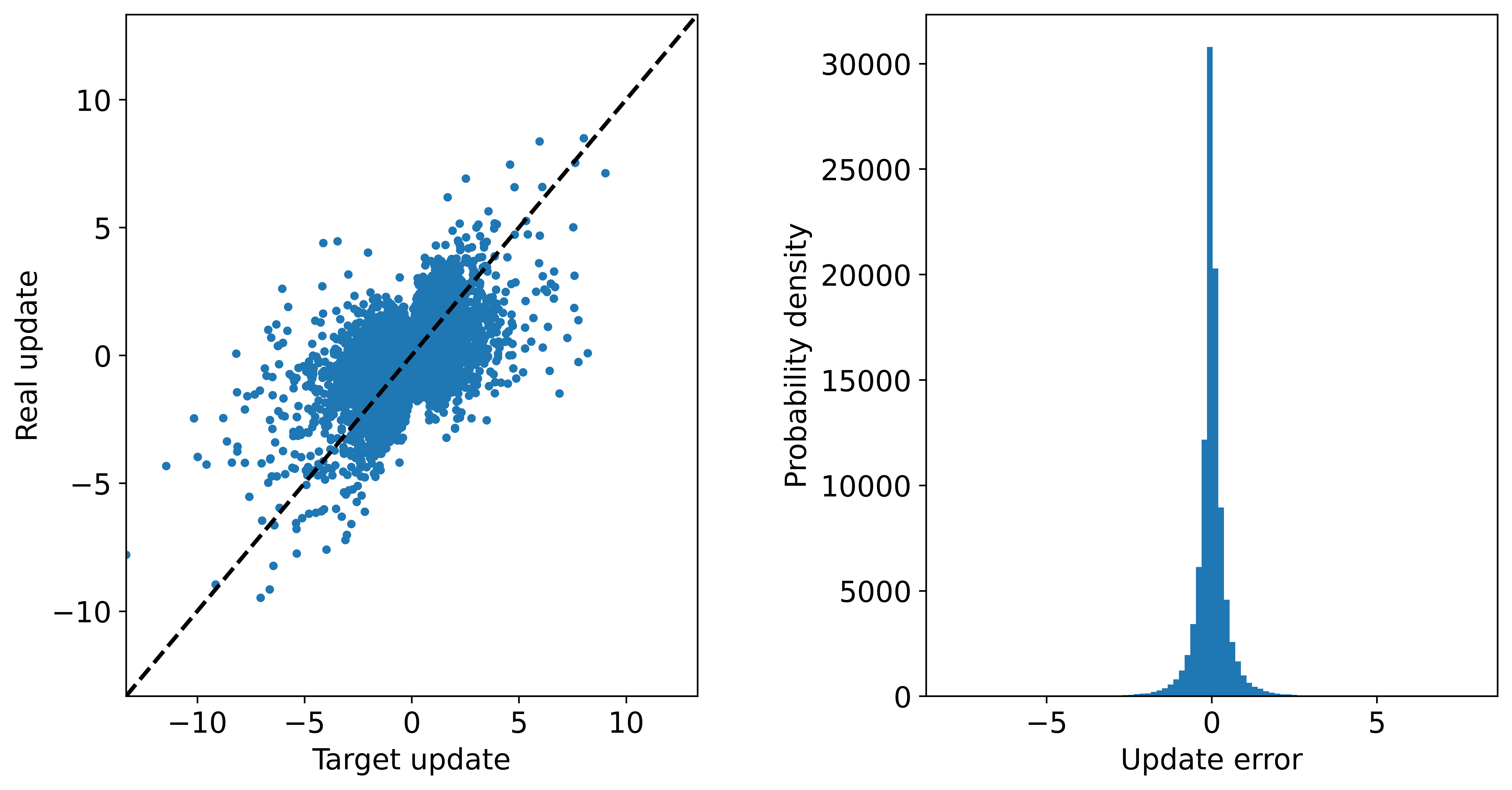}
    \caption{innercore update error Balanced DWMTJ SOT 300K standard}
    \label{Fig19}
\end{figure}
\vspace{10pt}
\vspace{10pt}
\begin{figure}[htbp]
    \centering
    \includegraphics[width=0.37\textwidth]{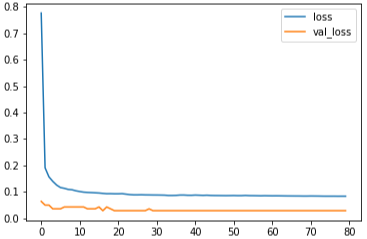}
    \caption{avg training loss vs avg validation loss}
    \label{Fig20}
\end{figure}
\begin{figure}[htbp]
    \centering
    \includegraphics[width=0.4\textwidth]{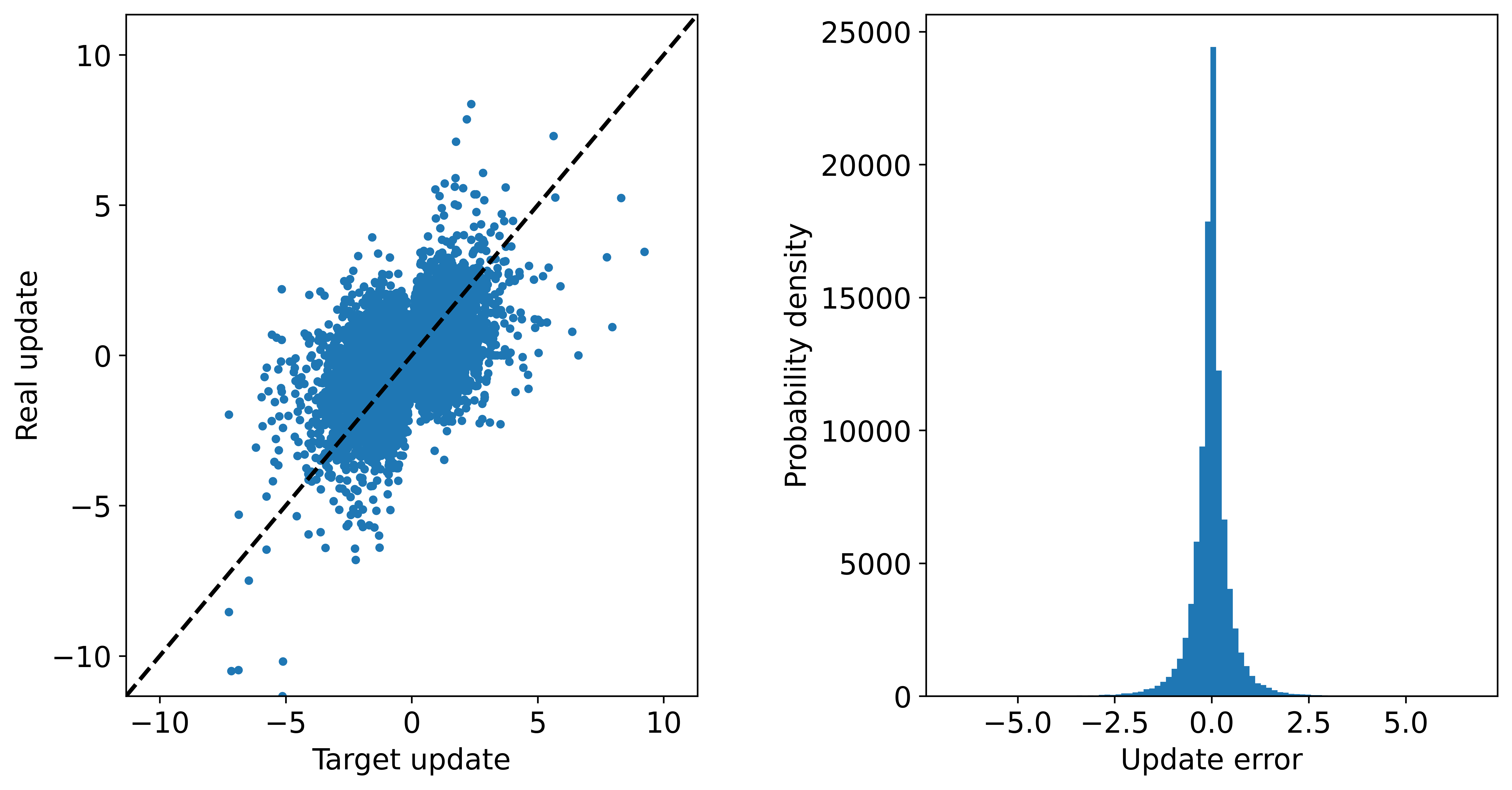}
    \caption{innercore update error Balanced DWMTJ SOT 300K multi}
    \label{Fig21}
\end{figure}
\begin{figure}[htbp]
    \centering
    \includegraphics[width=0.4\textwidth]{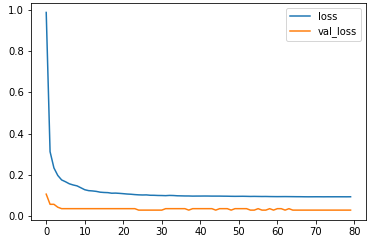}
    \caption{avg training loss vs avg validation loss}
    \label{Fig22}
\end{figure}

\subsection{DWMTJ SOT 400k}
\begin{figure}[htbp]
    \centering
    \includegraphics[width=0.4\textwidth]{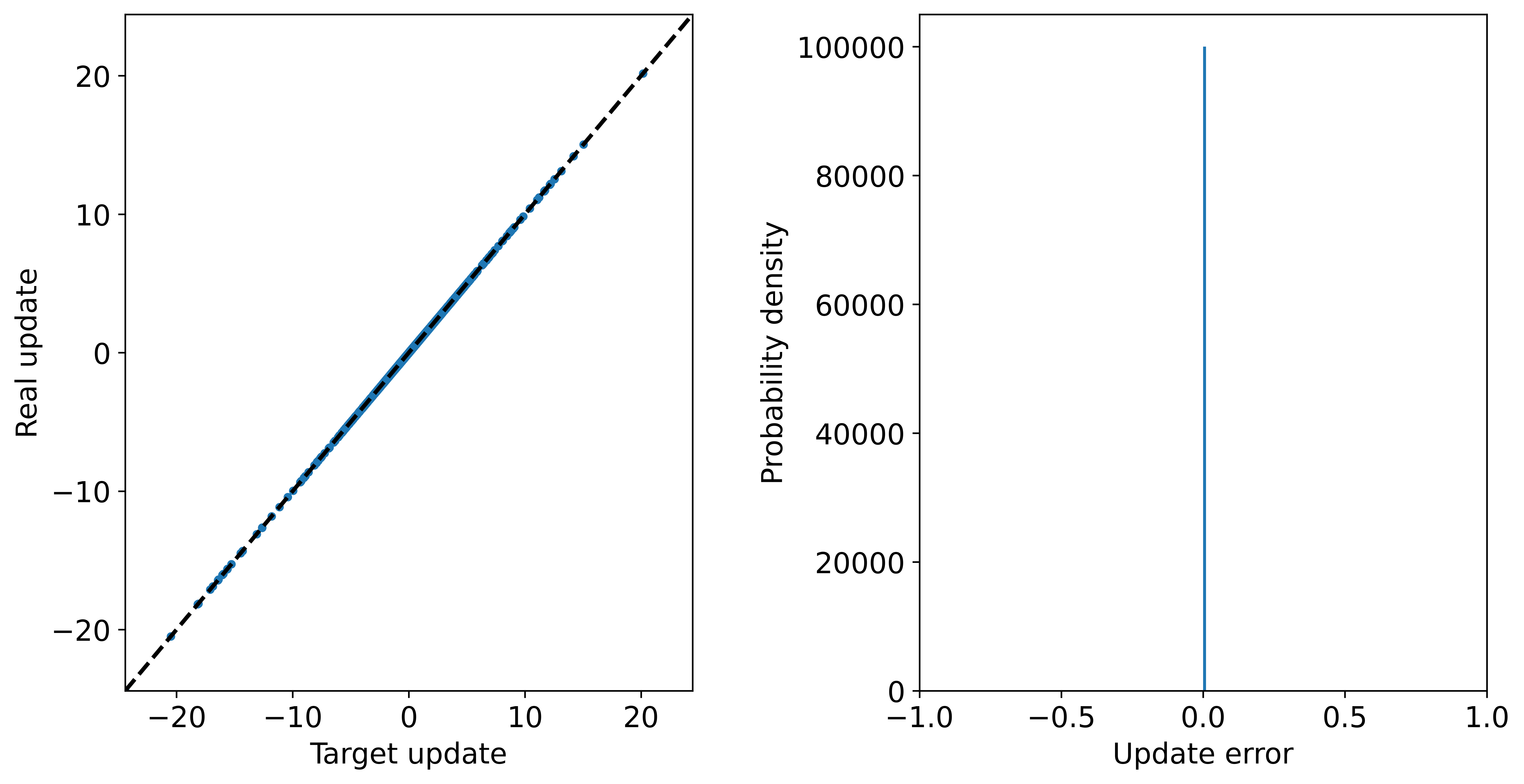}
    \caption{innercore update error Balanced DWMTJ SOT 400K numeric}
    \label{Fig23}
\end{figure}

\vspace{10pt}
\vspace{10pt}
\vspace{10pt}

\begin{figure}[htbp]
    \centering
    \includegraphics[width=0.35\textwidth]{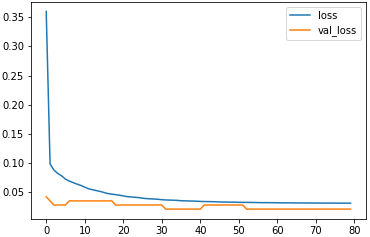}
    \caption{avg training loss vs avg validation loss}
    \label{Fig24}
\end{figure}

\begin{figure}[htbp]
    \centering
    \includegraphics[width=0.4\textwidth]{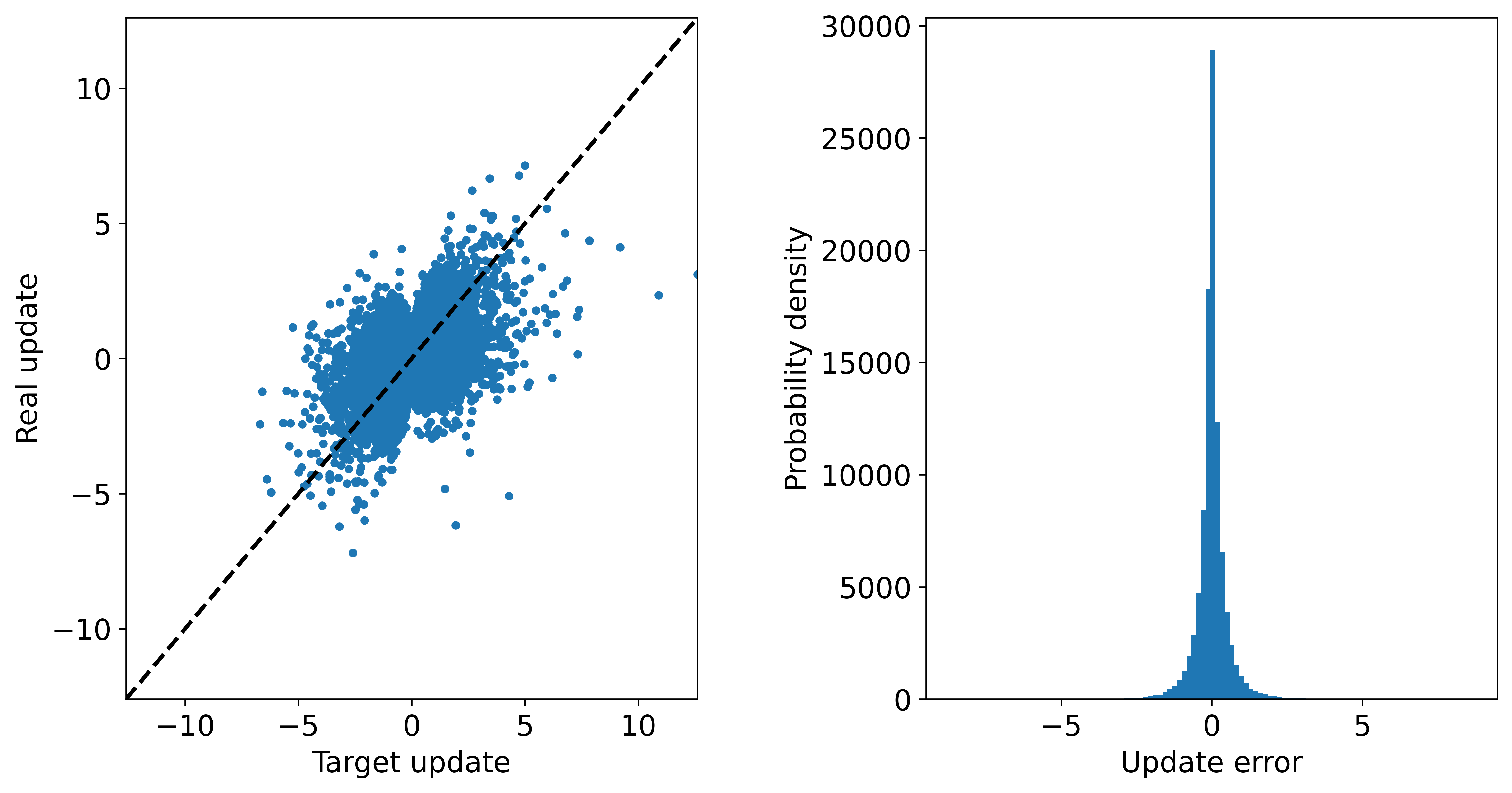}
    \caption{innercore update error Balanced DWMTJ SOT 400K standard}
    \label{Fig25}
\end{figure}

\begin{figure}[htbp]
    \centering
    \includegraphics[width=0.4\textwidth]{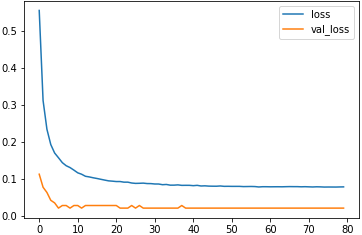}
    \caption{avg training loss vs avg validation loss}
    \label{Fig26}
\end{figure}

\begin{figure}[htbp]
    \centering
    \includegraphics[width=0.4\textwidth]{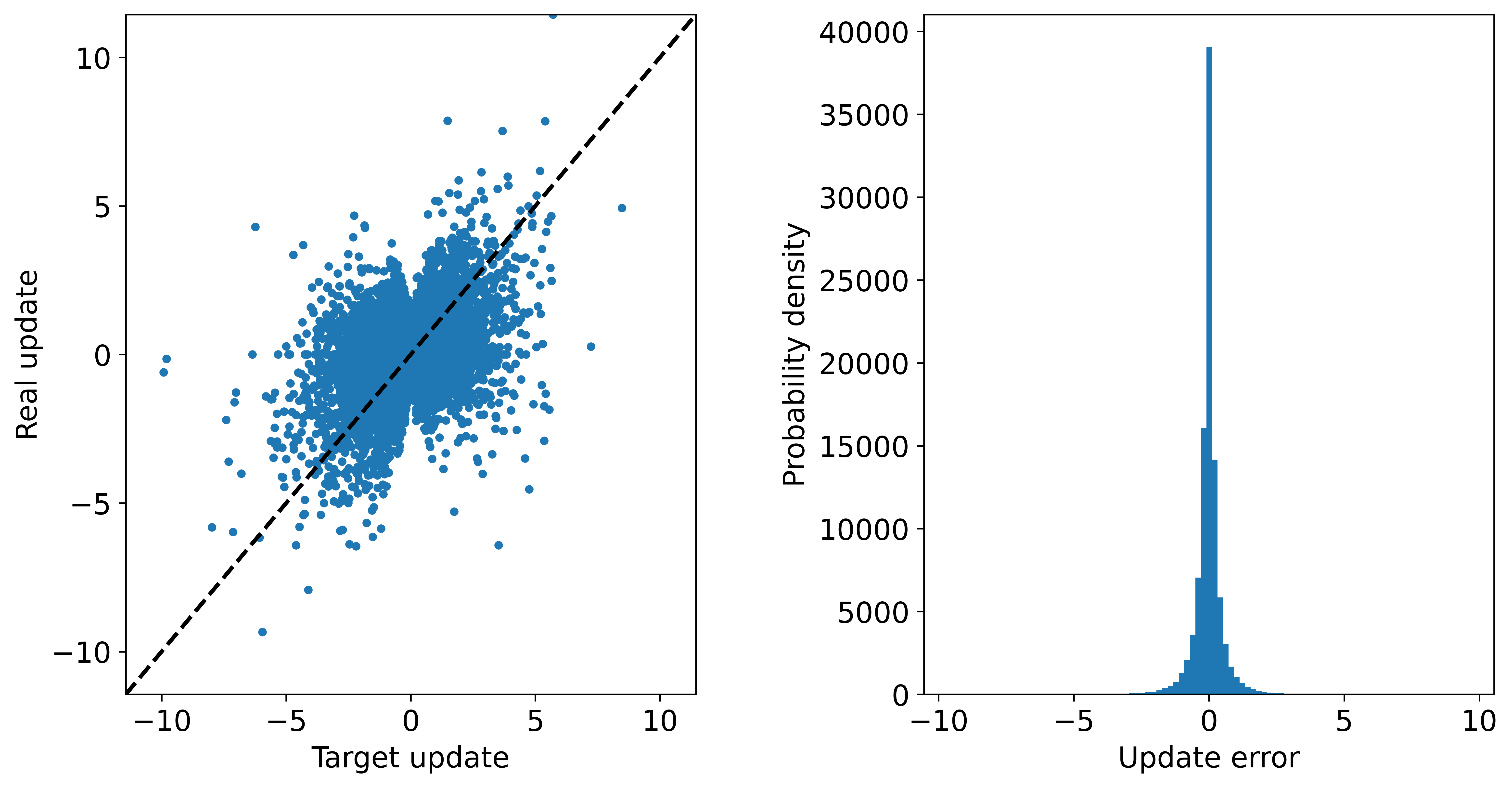}
    \caption{innercore update error Balanced DWMTJ SOT 400K multi}
    \label{Fig27}
\end{figure}

\begin{figure}[htbp]
    \centering
    \includegraphics[width=0.4\textwidth]{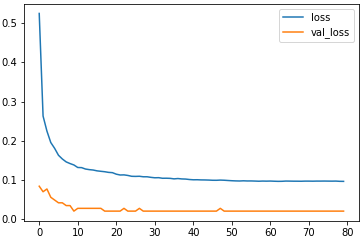}
    \caption{avg training loss vs avg validation loss}
    \label{Fig28}
\end{figure}

\vspace{10pt}
\vspace{10pt}
\vspace{10pt}

\subsection{DWMTJ STT 0k}
\begin{figure}[htbp]
    \centering
    \includegraphics[width=0.4\textwidth]{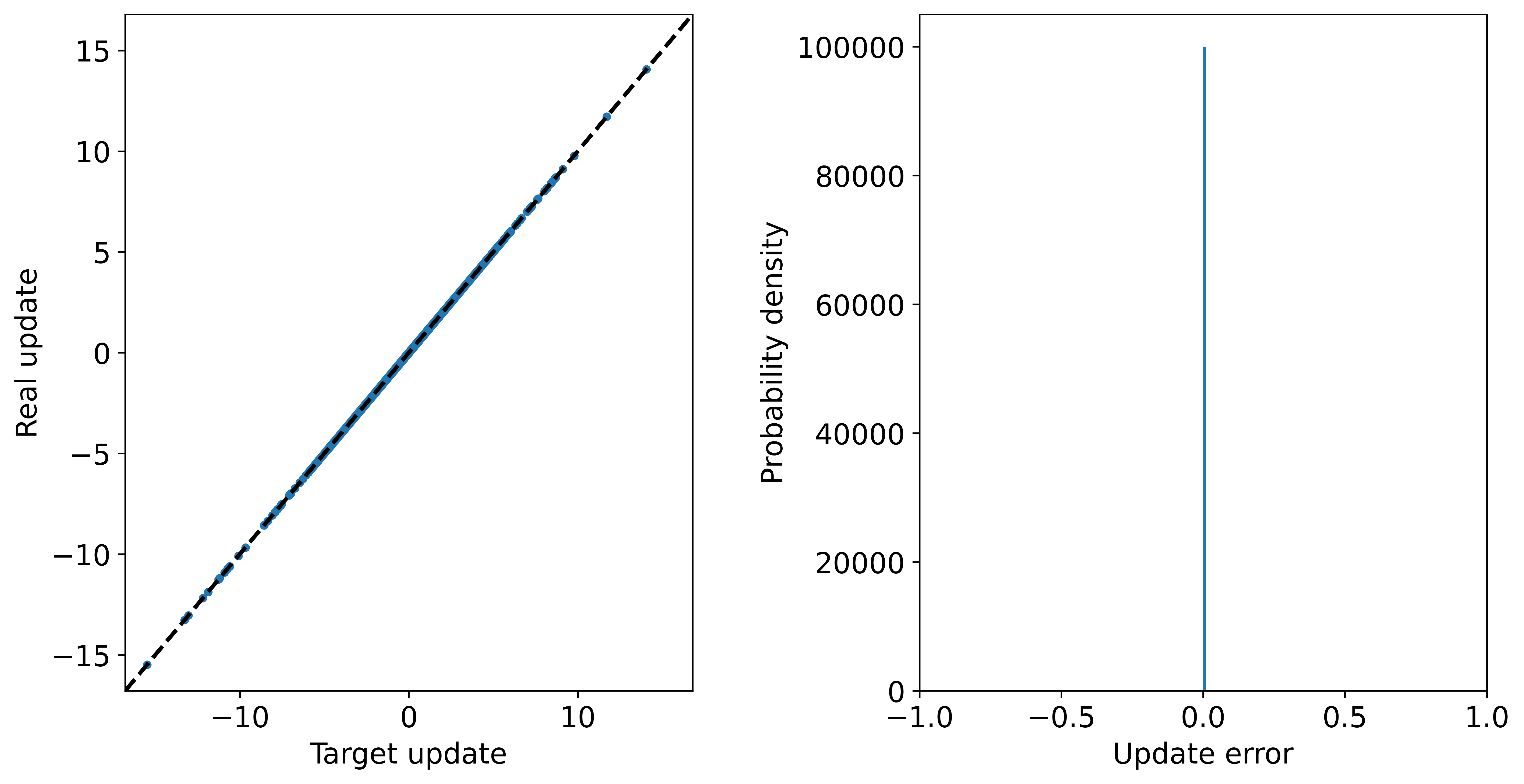}
    \caption{innercore update error Balanced DWMTJ STT 0K numeric}
    \label{Fig29}
\end{figure}

\begin{figure}[htbp]
    \centering
    \includegraphics[width=0.35\textwidth]{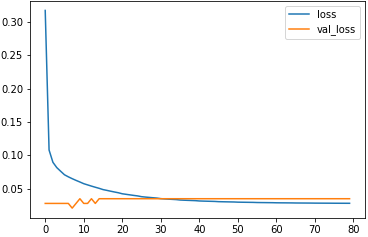}
    \caption{avg training loss vs avg validation loss}
    \label{Fig30}
\end{figure}

\begin{figure}[htbp]
    \centering
    \includegraphics[width=0.4\textwidth]{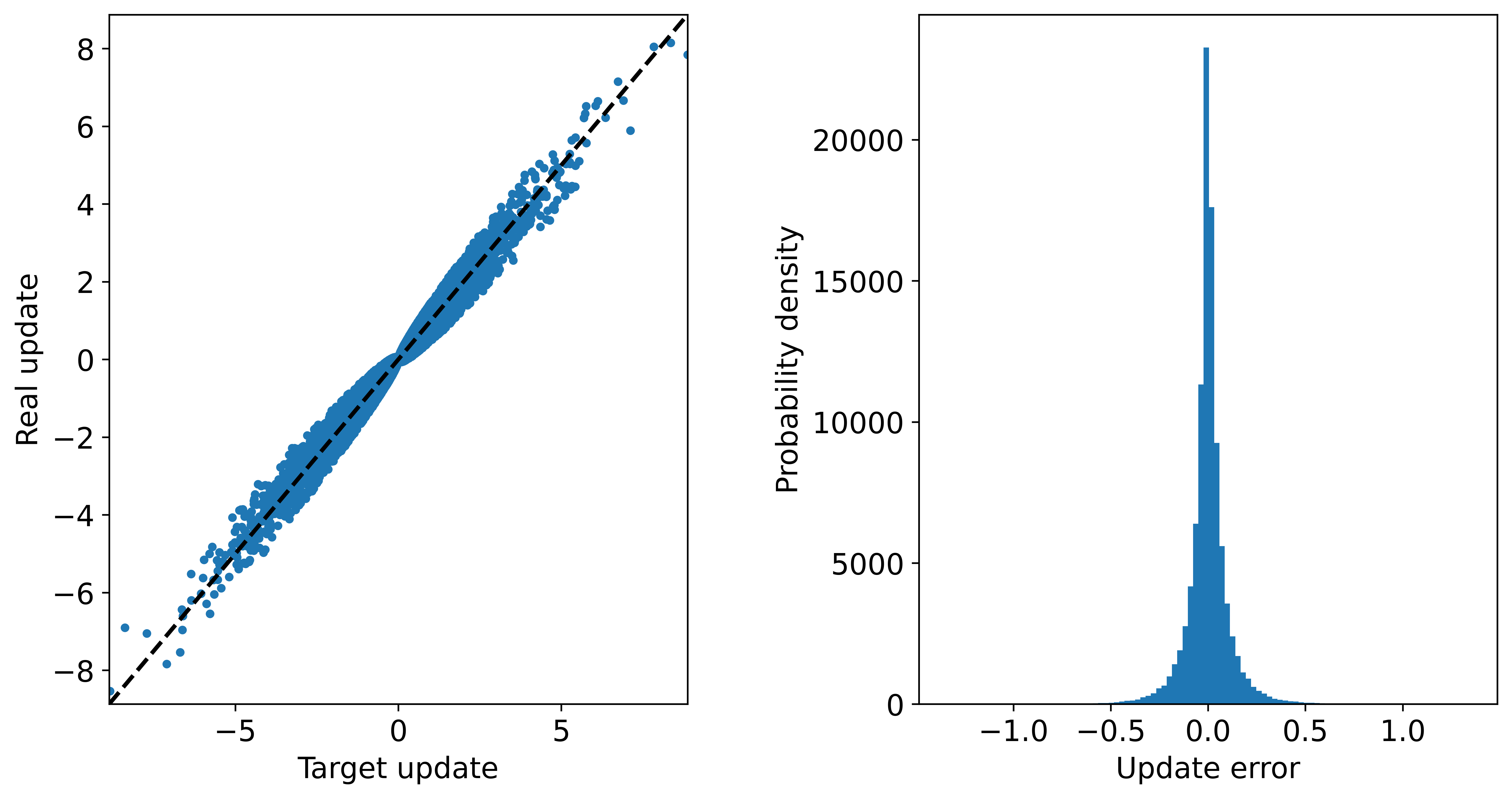}
    \caption{innercore update error Balanced DWMTJ STT 0K standard}
    \label{Fig31}
\end{figure}

\begin{figure}[htbp]
    \centering
    \includegraphics[width=0.37\textwidth]{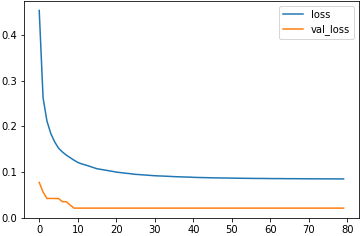}
    \caption{avg training loss vs avg validation loss}
    \label{Fig32}
\end{figure}

\vspace{30pt}
\vspace{30pt}
\vspace{30pt}

\begin{figure}[htbp]
    \centering
    \includegraphics[width=0.4\textwidth]{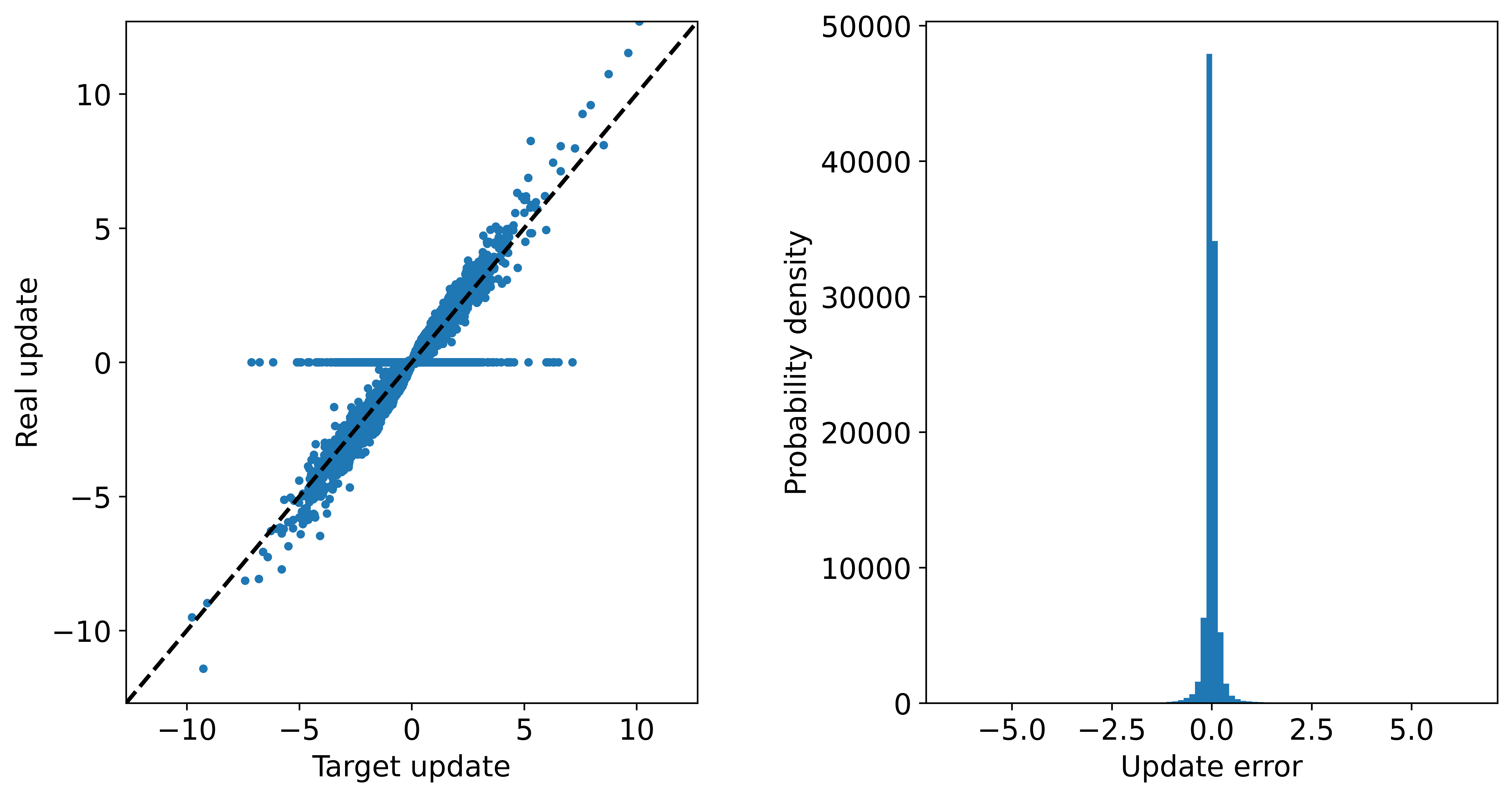}
    \caption{innercore update error Balanced DWMTJ STT 0K multi}
    \label{Fig33}
\end{figure}

\begin{figure}[htbp]
    \centering
    \includegraphics[width=0.4\textwidth]{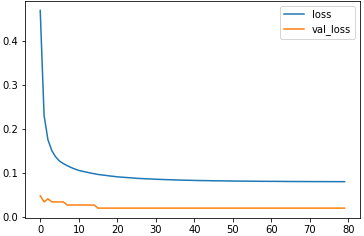}
    \caption{avg training loss vs avg validation loss}
    \label{Fig34}
\end{figure}

\subsection{DWMTJ STT 300k}
\begin{figure}[htbp]
    \centering
    \includegraphics[width=0.37\textwidth]{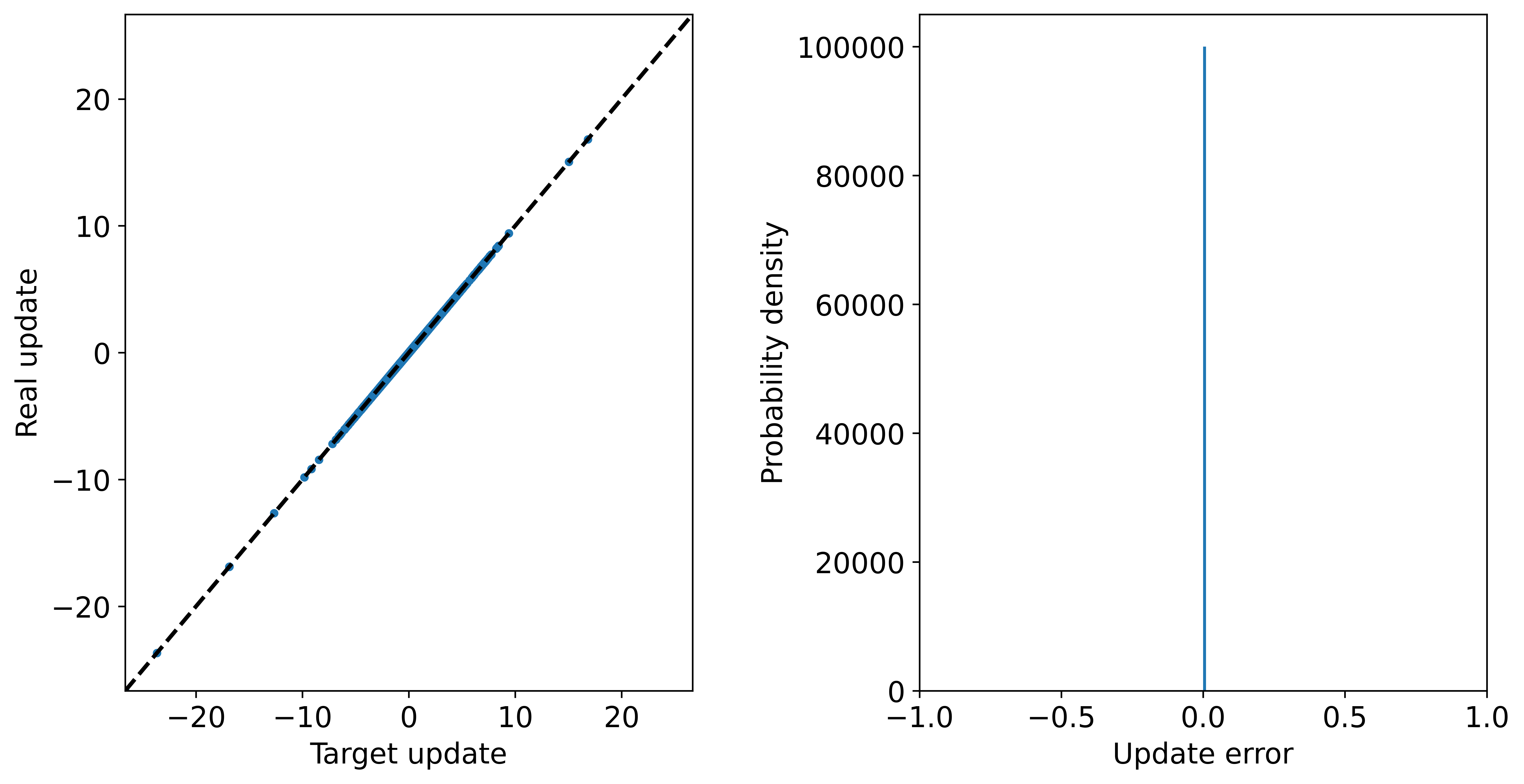}
    \caption{innercore update error Balanced DWMTJ STT 300K numeric}
    \label{Fig35}
\end{figure}

\begin{figure}[htbp]
    \centering
    \includegraphics[width=0.36\textwidth]{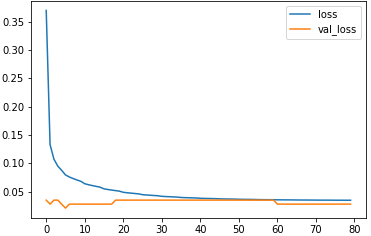}
    \caption{avg training loss vs avg validation loss}
    \label{Fig36}
\end{figure}

\begin{figure}[htbp]
    \centering
    \includegraphics[width=0.4\textwidth]{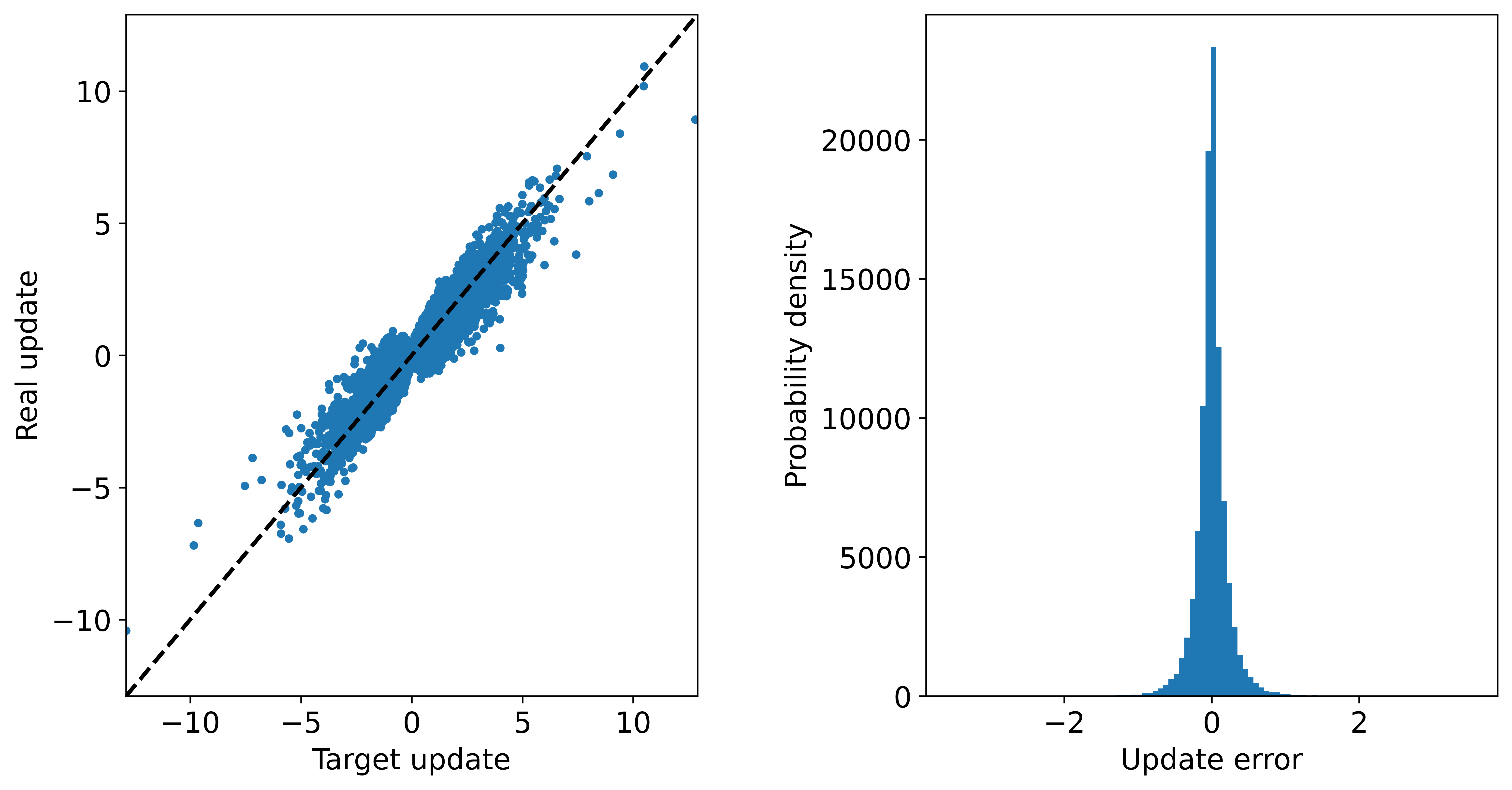}
    \caption{innercore update error Balanced DWMTJ STT 300K standard}
    \label{Fig37}
\end{figure}

\begin{figure}[htbp]
    \centering
    \includegraphics[width=0.4\textwidth]{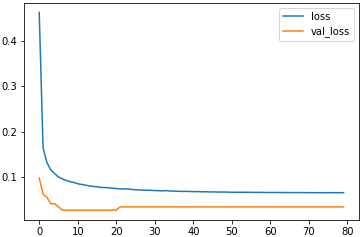}
    \caption{avg training loss vs avg validation loss}
    \label{Fig38}
\end{figure}

\begin{figure}[htbp]
    \centering
    \includegraphics[width=0.4\textwidth]{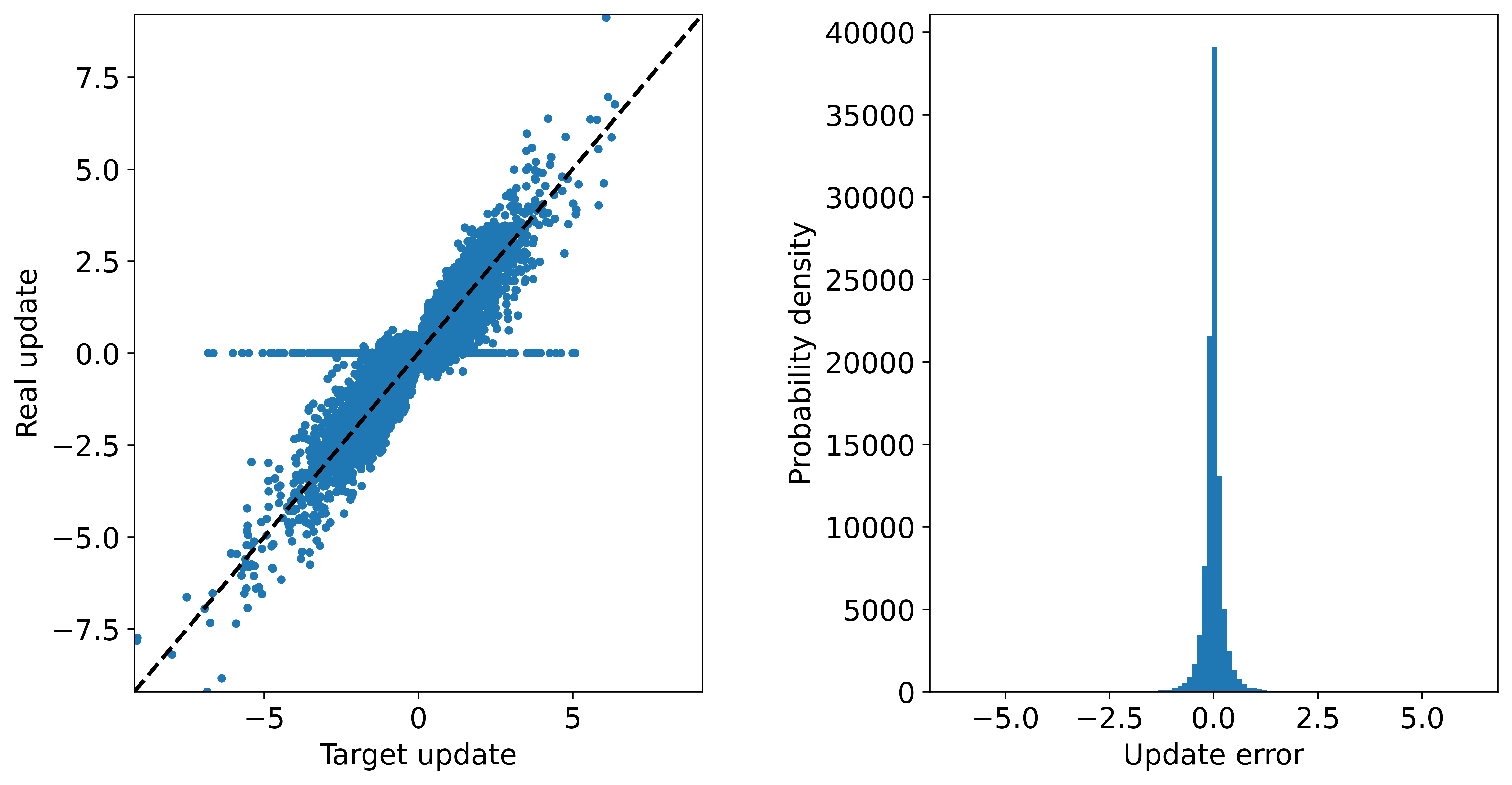}
    \caption{innercore update error Balanced DWMTJ STT 300K multi}
    \label{Fig39}
\end{figure}

\begin{figure}[htbp]
    \centering
    \includegraphics[width=0.4\textwidth]{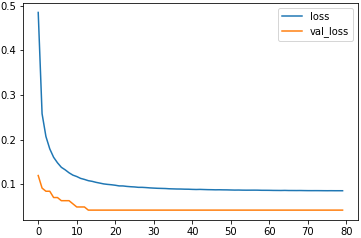}
    \caption{avg training loss vs avg validation loss}
    \label{Fig40}
\end{figure}

\vspace{30pt}
\vspace{30pt}
\vspace{30pt}

\subsection{DWMTJ STT 400k}
\begin{figure}[htbp]
    \centering
    \includegraphics[width=0.37\textwidth]{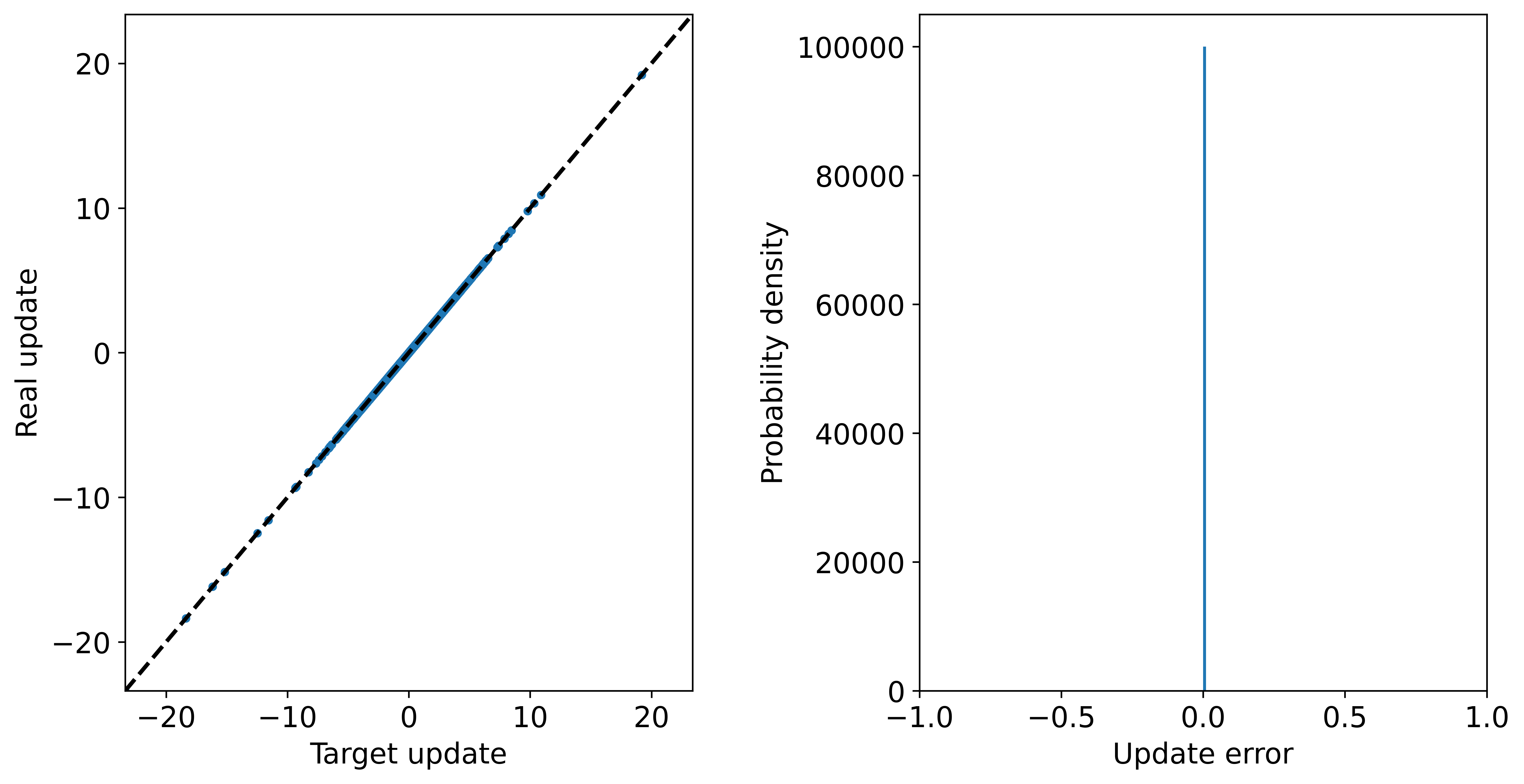}
    \caption{innercore update error Balanced DWMTJ STT 400K numeric}
    \label{Fig41}
\end{figure}

\begin{figure}[htbp]
    \centering
    \includegraphics[width=0.4\textwidth]{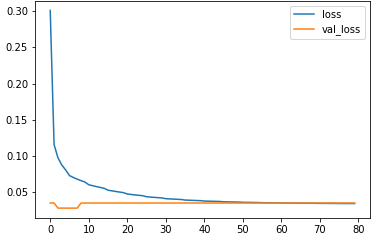}
    \caption{avg training loss vs avg validation loss}
    \label{Fig42}
\end{figure}

\begin{figure}[htbp]
    \centering
    \includegraphics[width=0.37\textwidth]{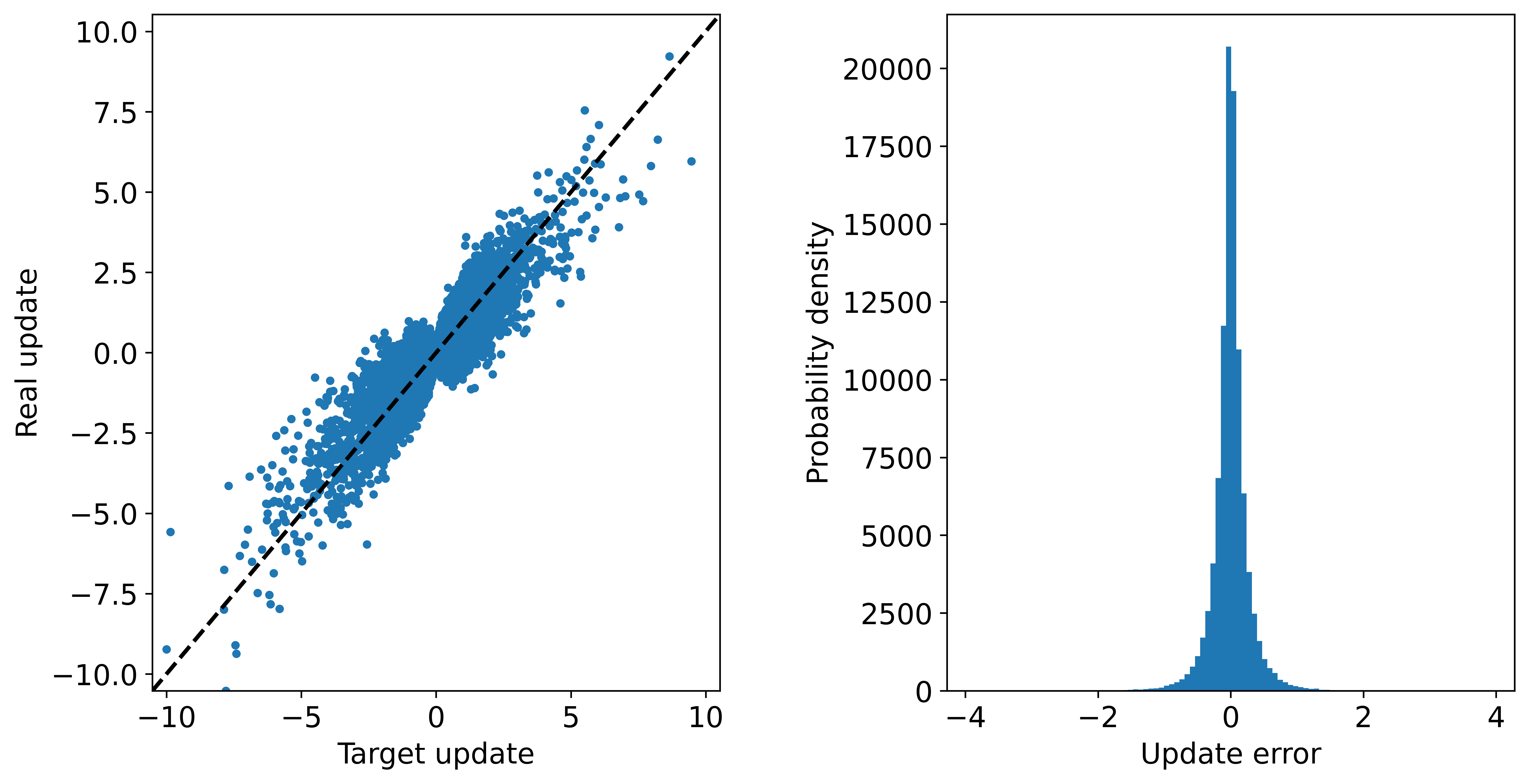}
    \caption{innercore update error Balanced DWMTJ STT 400K standard}
    \label{Fig43}
\end{figure}

\begin{figure}[htbp]
    \centering
    \includegraphics[width=0.4\textwidth]{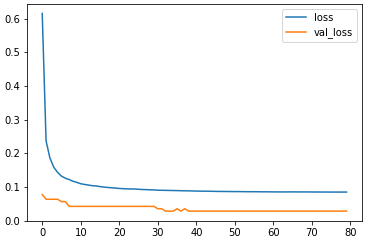}
    \caption{avg training loss vs avg validation loss}
    \label{Fig44}
\end{figure}

\begin{figure}[htbp]
    \centering
    \includegraphics[width=0.4\textwidth]{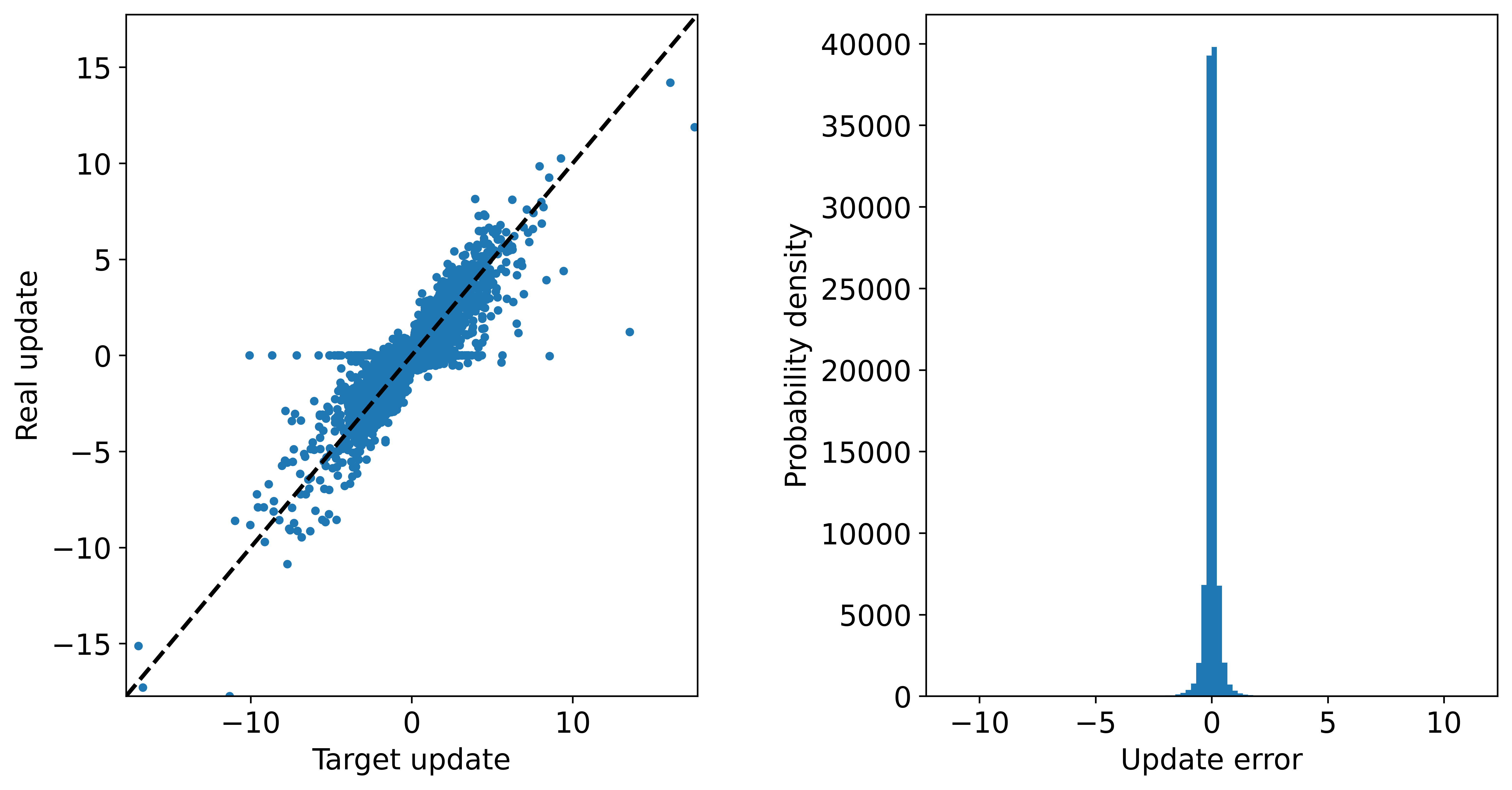}
    \caption{innercore update error Balanced DWMTJ STT 400K multi}
    \label{Fig45}
\end{figure}

\begin{figure}[htbp]
    \centering
    \includegraphics[width=0.4\textwidth]{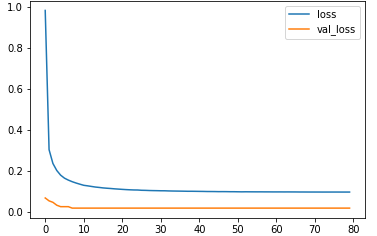}
    \caption{avg training loss vs avg validation loss}
    \label{Fig46}
\end{figure}

\vspace{30pt}
\vspace{30pt}
\vspace{30pt}

\subsection{ENODe}
\begin{figure}[htbp]
    \centering
    \includegraphics[width=0.39\textwidth]{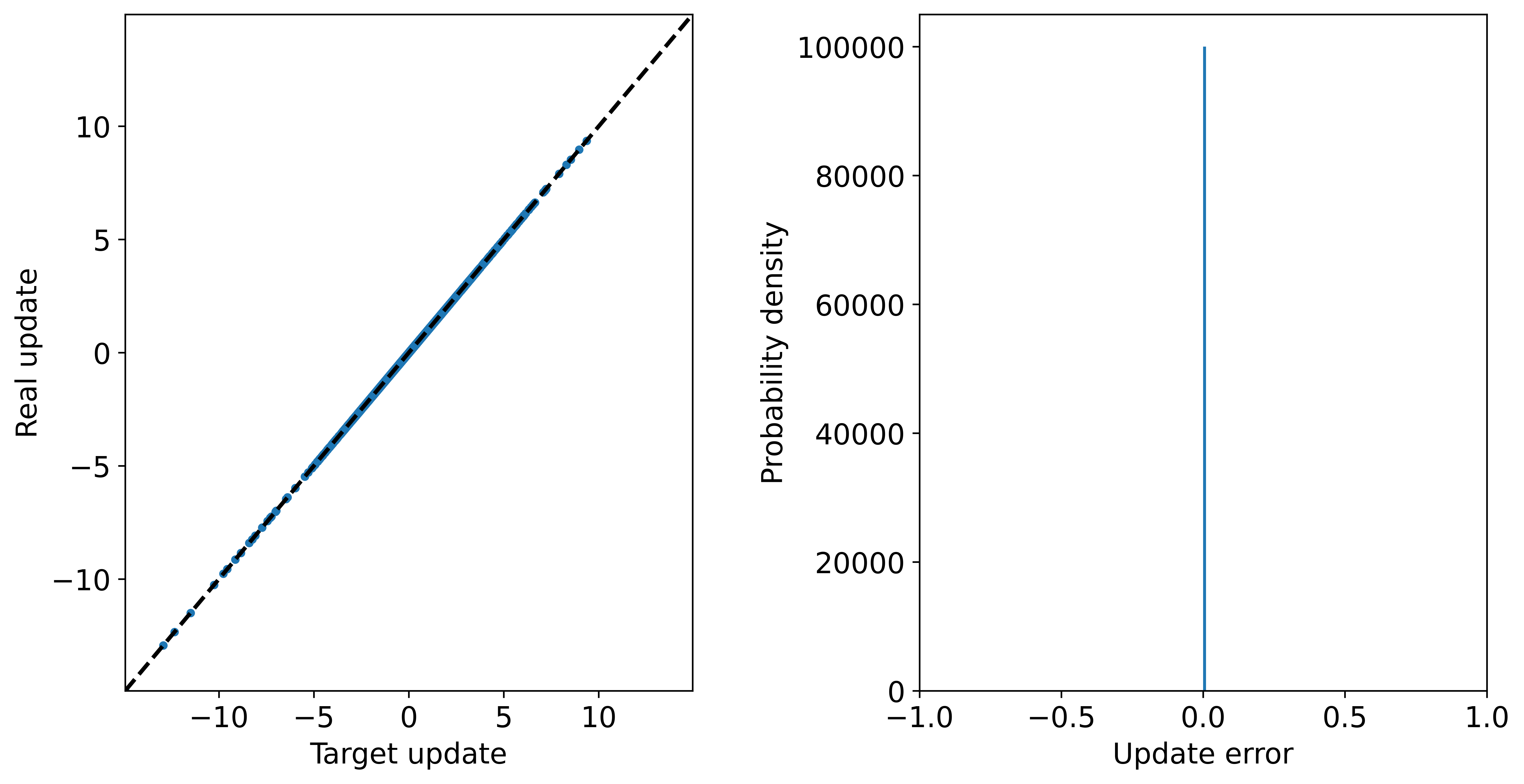}
    \caption{innercore update error Balanced ENODe numeric}
    \label{Fig47}
\end{figure}

\begin{figure}[htbp]
    \centering
    \includegraphics[width=0.39\textwidth]{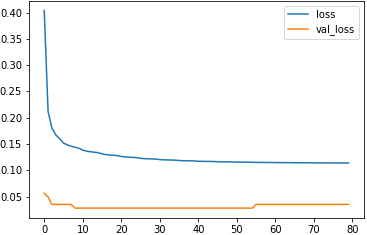}
    \caption{avg training loss vs avg validation loss}
    \label{Fig48}
\end{figure}

\begin{figure}[htbp]
    \centering
    \includegraphics[width=0.4\textwidth]{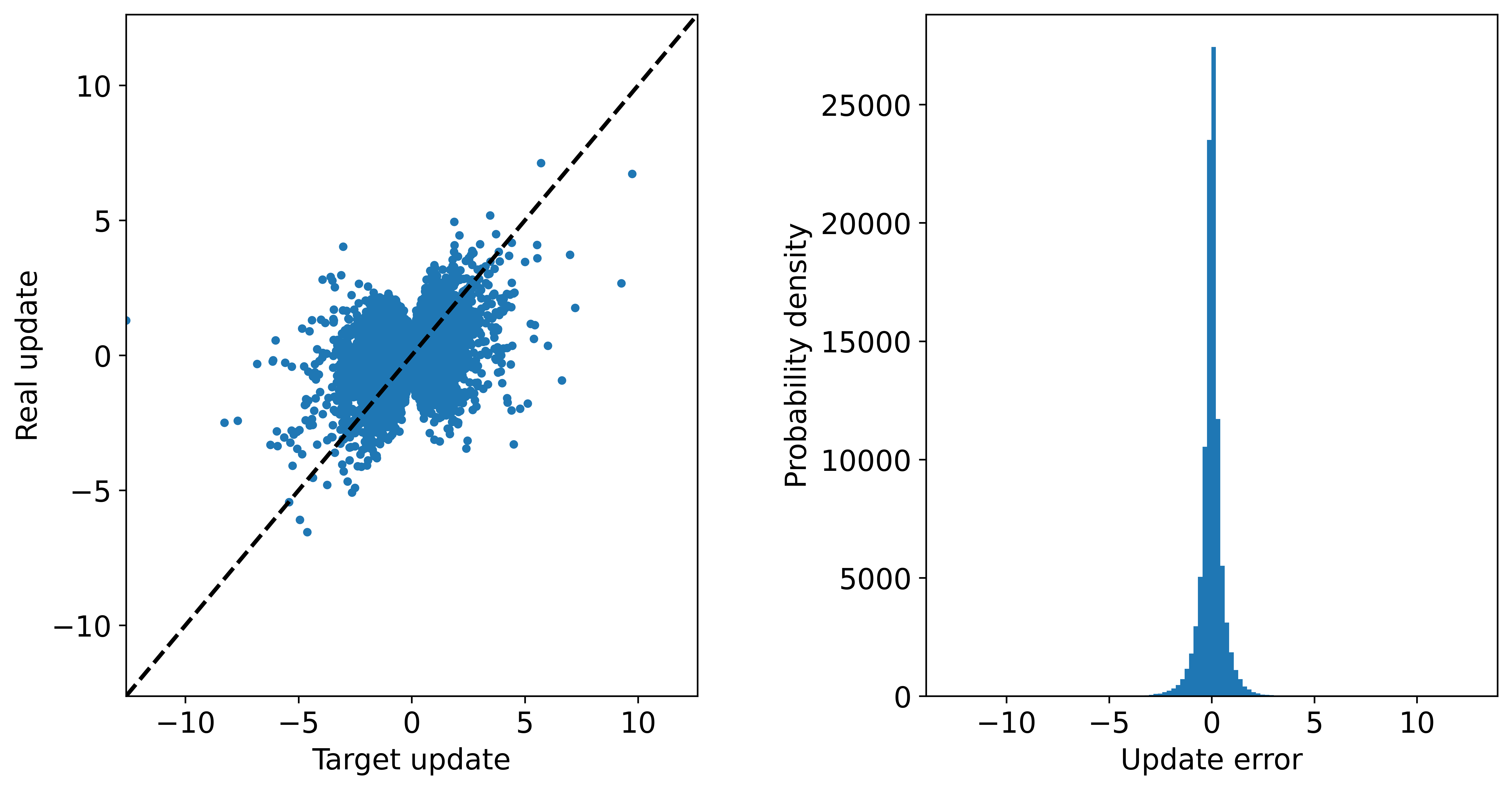}
    \caption{innercore update error Balanced ENODe standard}
    \label{Fig49}
\end{figure}

\begin{figure}[htbp]
    \centering
    \includegraphics[width=0.4\textwidth]{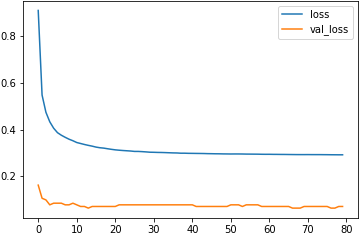}
    \caption{avg training loss vs avg validation loss}
    \label{Fig50}
\end{figure}

\begin{figure}[htbp]
    \centering
    \includegraphics[width=0.4\textwidth]{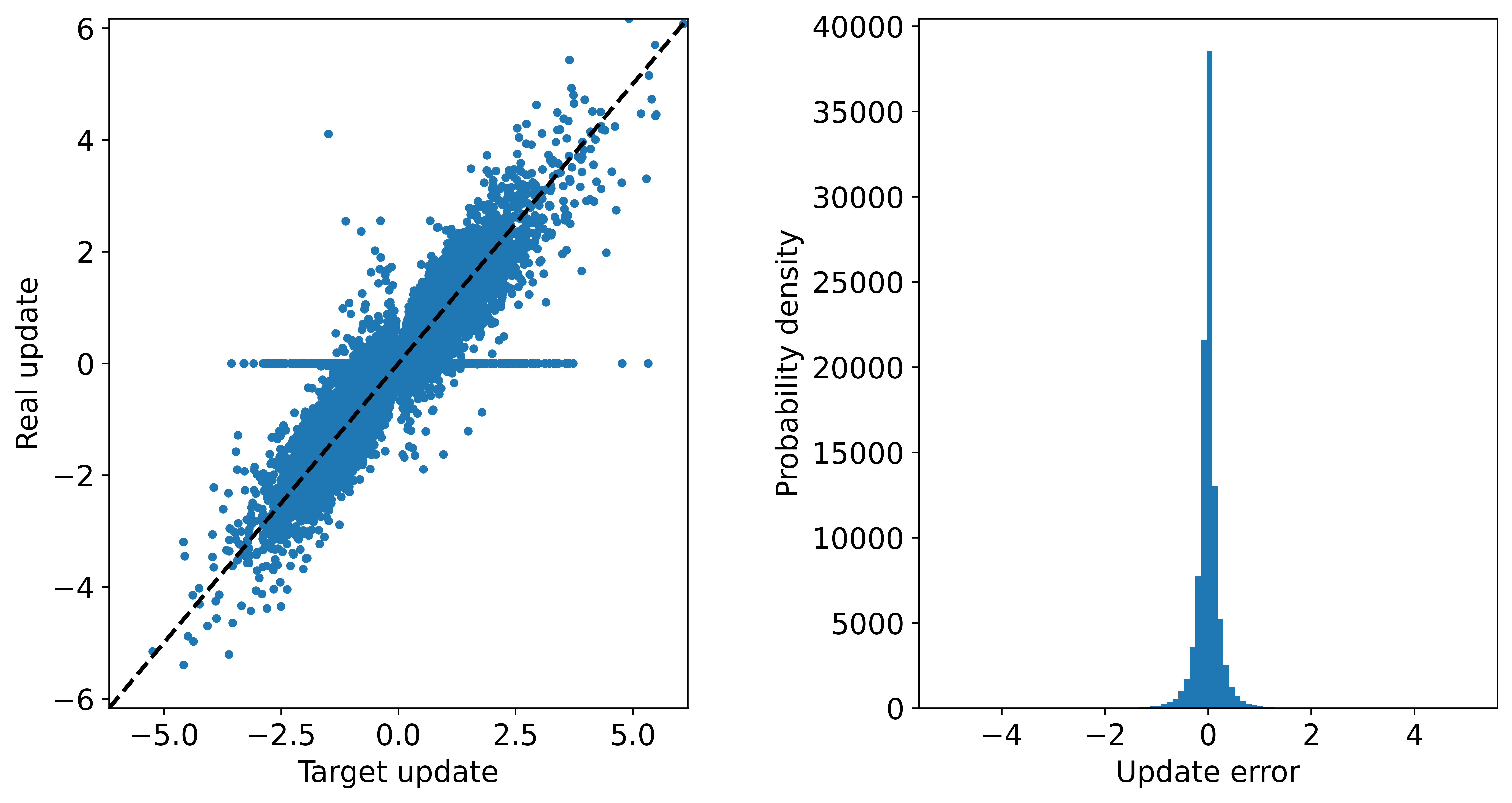}
    \caption{innercore update error Balanced ENODe multi}
    \label{Fig51}
\end{figure}

\begin{figure}[htbp]
    \centering
    \includegraphics[width=0.4\textwidth]{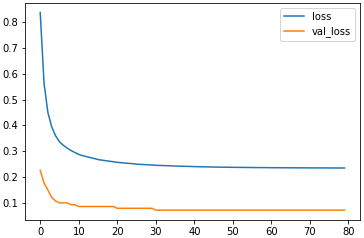}
    \caption{avg training loss vs avg validation loss}
    \label{Fig52}
\end{figure}

\vspace{30pt}
\vspace{30pt}
\vspace{30pt}

\subsection{TaOx}
\begin{figure}[htbp]
    \centering
    \includegraphics[width=0.39\textwidth]{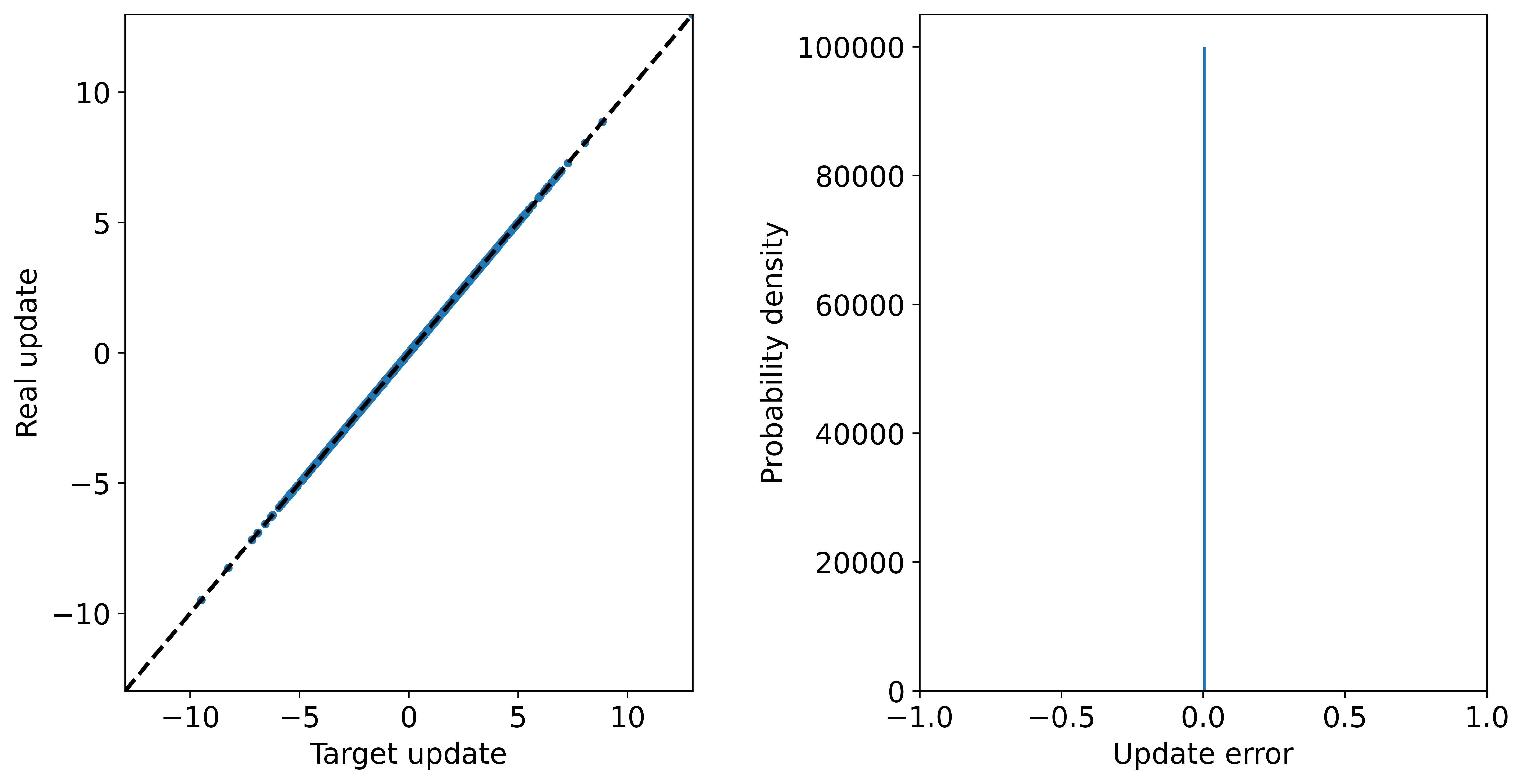}
    \caption{innercore update error Balanced TaOx numeric}
    \label{Fig53}
\end{figure}

\begin{figure}[htbp]
    \centering
    \includegraphics[width=0.39\textwidth]{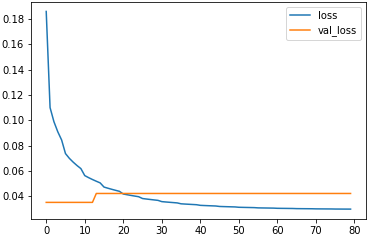}
    \caption{avg training loss vs avg validation loss}
    \label{Fig54}
\end{figure}

\begin{figure}[htbp]
    \centering
    \includegraphics[width=0.39\textwidth]{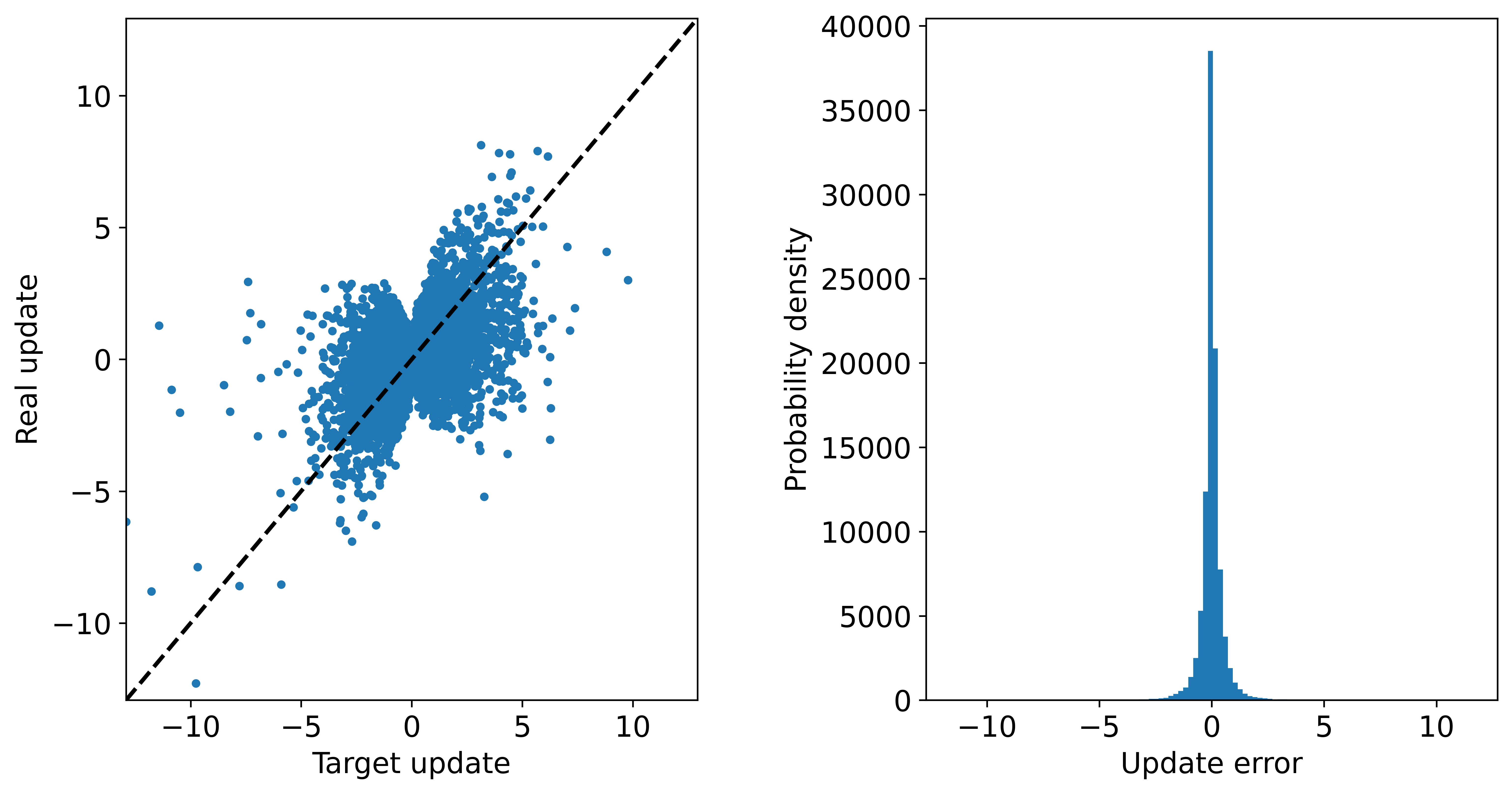}
    \caption{innercore update error Balanced TaOx standard}
    \label{Fig55}
\end{figure}

\begin{figure}[htbp]
    \centering
    \includegraphics[width=0.36\textwidth]{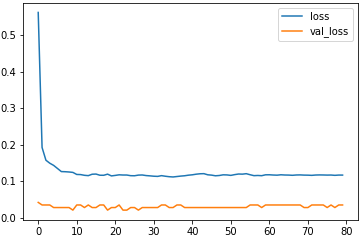}
    \caption{avg training loss vs avg validation loss}
    \label{Fig56}
\end{figure}

\vspace{30pt}
\vspace{30pt}
\vspace{30pt}

\begin{figure}[htbp]
    \centering
    \includegraphics[width=0.4\textwidth]{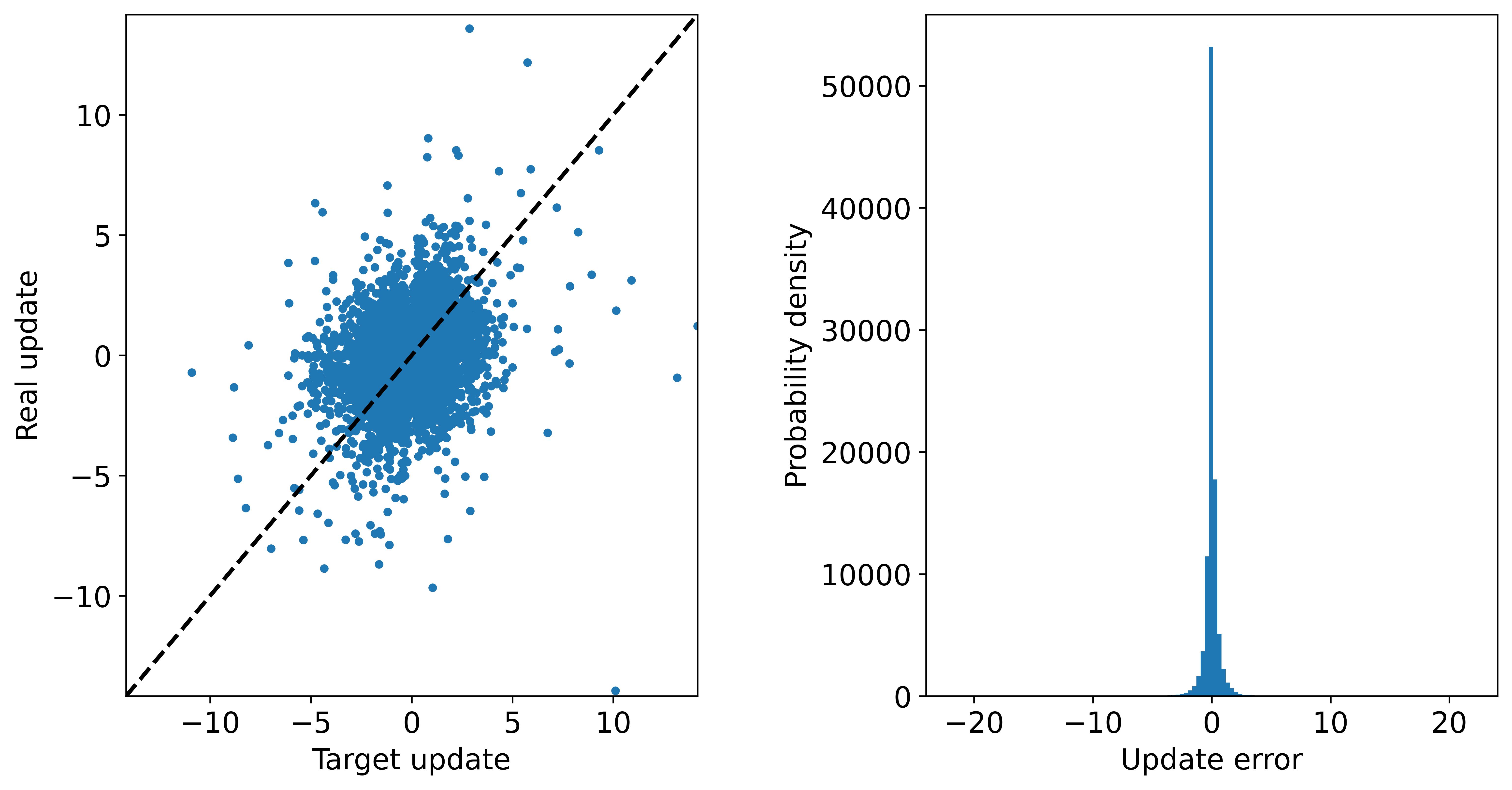}
    \caption{innercore update error Balanced TaOx multi}
    \label{Fig57}
\end{figure}
\begin{figure}[htbp]
    \centering
    \includegraphics[width=0.4\textwidth]{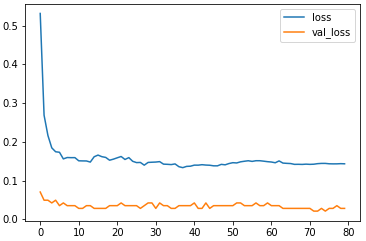}
    \caption{avg training loss vs avg validation loss}
    \label{Fig58}
\end{figure}
\bibliographystyle{IEEEtran}
\bibliography{references}
\end{document}